\documentclass[12pt,twoside]{article}
\usepackage{amssymb,amsmath,epsfig,color}
\usepackage{pstricks}

\definecolor{asblue}{rgb}{0,0,0.6}
\definecolor{asred}{rgb}{0.8,0,0}
\definecolor{asgreen}{rgb}{0,0.6,0}
\definecolor{asorange}{rgb}{1.0,0.5,0.3}
\newrgbcolor{ascyan}{0 0.6 0.7}
\definecolor{orange}{rgb}{0.7,0.3,0.1}
\newrgbcolor{aspurple}{0.5 0.1 0.5}
\newrgbcolor{asbrown}{0.7 0.4 0.2}

\newcommand{\Funct}[2]{#1\!\left(#2\right)}

\newcommand{\Parth}[1]{\left( #1 \right)}
\newcommand{\Floor}[1]{\left\lfloor #1 \right\rfloor}

\newcommand{\Inv}[1]{\frac{1}{#1}}

\newcommand{\Integers}{\mathbb{Z}}

\newcommand{\InfSum}[1]{\sum_{#1}^{\infty}}
\newcommand{\DISum}[1]{\sum_{#1=-\infty}^{\infty}}
\newcommand{\DIInt}[1]{\int_{-\infty}^{\infty} #1 \: \mathrm{d}}

\newcommand{\Sinc}[1]{\mbox{sinc}\!\left(#1\right)}

\newcommand{\FT}[1]{#1(e^{j \omega})}
\newcommand{\FTf}[2]{#1(e^{j 2 \pi {#2}})}

\newcommand{\Cne}[1]{e^{-j 2 \pi #1}}

\begin{document}

\begin{titlepage}

%
\begin{figure}[tp]
\setlength{\unitlength}{1mm}
\begin{picture}(160,50)
\put(140,20){\includegraphics[width=20mm]{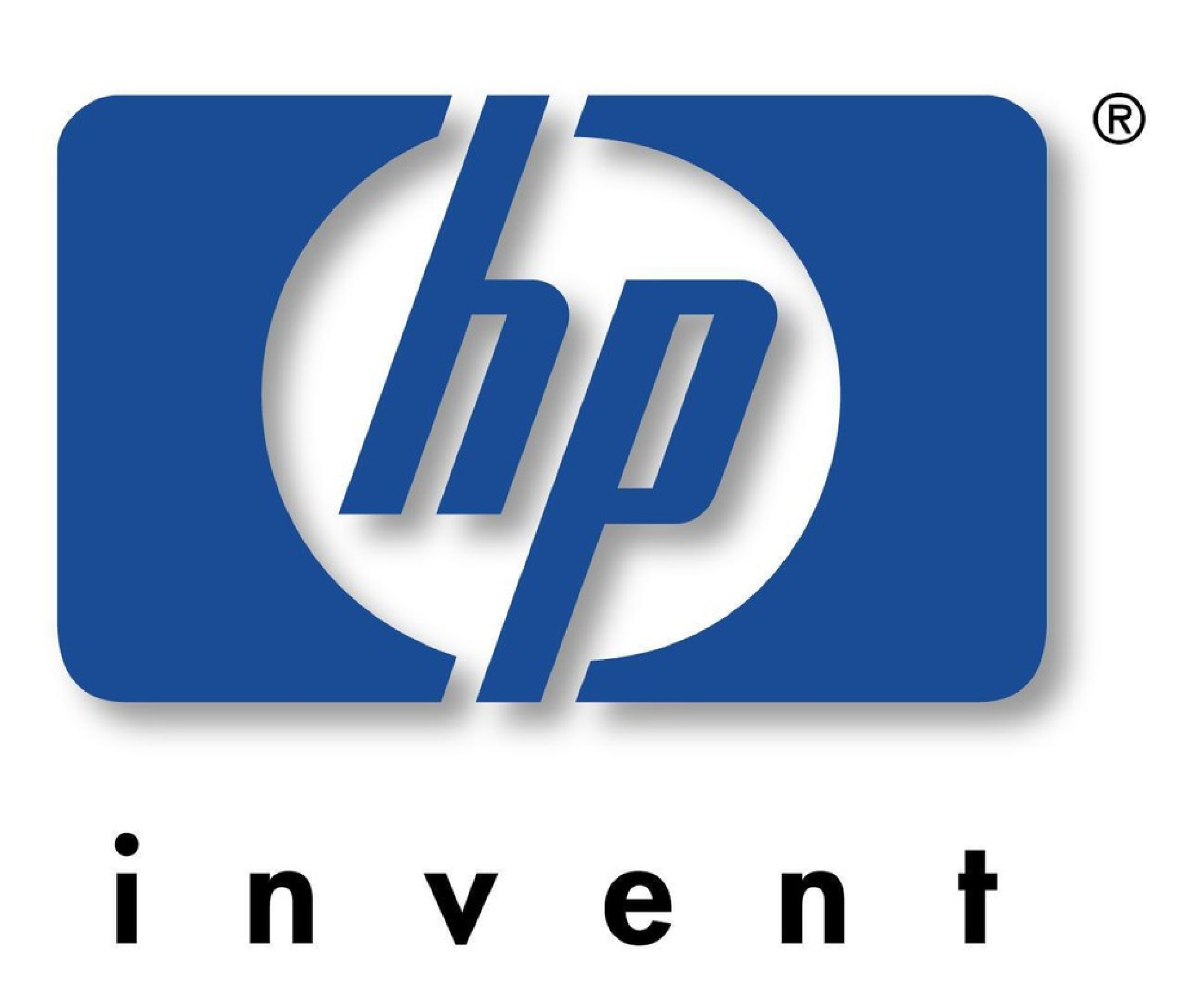}}
\end{picture}
\end{figure}

\renewcommand*{\thefootnote}{\fnsymbol{footnote}}

\noindent {\Large \bf New Filters for Image Interpolation and Resizing} \\

\vspace{2mm}
\noindent {\bf Amir Said}\\
Media Technologies Laboratory\\
Hewlett-Packard Laboratories, Palo Alto, CA, USA\\
\ \\
HPL-2007-179\\
Nov. 2, 2007\footnote{Internal Accession Data Only, Approved for External Publication\\This report contains material published in: A. Said, ``A new class of filters for image interpolation and resizing,'' {\em Proc. IEEE Int. Conf. Image Processing,} 2007.}

\vspace{10mm}

\noindent \parbox[t]{40mm}{\noindent \small interpolation kernels,\\image processing,\\image resizing}
\parbox[t]{123mm}{\small We
propose a new class of kernels to simplify the design of filters for image interpolation and resizing. Their properties
are defined according to two parameters, specifying the width of the transition band and the height of a unique
sidelobe. By varying these parameters it is possible to efficiently explore the space with only the filters that are
suitable for image interpolation and resizing, and identify the filter that is best for a given application. These two
parameters are also sufficient to obtain very good approximations of many commonly-used interpolation
kernels. We also show that, because the Fourier transforms of these kernels have very fast decay, these filters
produce better results when time-stretched for image downsizing.}

\end{titlepage}

\thispagestyle{empty}
\clearpage


\setcounter{page}{1}
\renewcommand*{\thefootnote}{\arabic{footnote}}

\title{\bf New Filters for Image Interpolation and Resizing}

\author{Amir Said \\ Hewlett Packard Laboratories, Palo Alto, CA}

\date{}

\maketitle

\begin{abstract}
We propose a new class of kernels to simplify the design of filters for image interpolation and resizing. Their
properties are defined according to two parameters, specifying the width of the transition band and the height of
a unique sidelobe. By varying these parameters it is possible to efficiently explore the space with only the
filters that are suitable for image interpolation and resizing, and identify the filter that is best for a given
application. These two parameters are also sufficient to obtain very good approximations of many commonly-used
interpolation kernels. We also show that, because the Fourier transforms of these kernels have very fast decay,
these filters produce better results when time-stretched for image downsizing.
\end{abstract}

\section{Introduction}\label{ssIntro}

Image resizing and interpolation (e.g., for rotation) are two of the most useful image processing operations, and
consequently there is a great amount of literature on the subject~\cite{Harris78}--\cite{Meijering02}. However, many
imaging professionals find the task of sorting out and implementing the most appropriate method quite challenging,
due to the great number of possibilities and conflicting opinions. It is common to settle for some very simple
approaches which were once meant to reduce complexity, or adopt one type that was shown to be excellent for one
application, without knowing that it may be suboptimal for other applications.

For instance, even in commercial products we find the mistake of using interpolation kernels for downsampling without
the necessary lowpass filtering. A less serious, but also common mistake, is to use for downsampling low-order filters
which can be good for interpolation, but have much worse properties when time-scaled for downsizing.

What is still missing is an approach that is more {\em convenient and easy to use,} with less emphasis on computational
complexity, and that yields high image quality. For that purpose we propose a family of parametrized functions that
are simple, and are designed with enough versatility so that well-known kernels can be very closely approximated by
simply using the proper parameters. This way it is easy to experiment and identify the parameters that are best for a certain type of image, without having to understand and implement several methods.

This paper is organized by first having a very brief review of sampling and interpolation concepts, which allow us
to establish the notation. Next, we define the proposed family of functions, and some of their basic properties. We
present a set of features that are desirable for interpolation and resizing kernels, and explain how well the proposed
functions support those features. Finally, we present some results of finding the approximation to some commonly-used
kernels, show how well these kernels are approximated, and discuss some additional features of the proposed kernels.

\section{Signal Sampling and Reconstruction}\label{ssSampling}

%
\begin{figure}[tb]
\centering
\setlength{\unitlength}{0.8mm}
\psset{unit=0.8mm}

\begin{pspicture}(120, 70)
\psset{linewidth=1.75pt,linecolor=asblue}
\psbezier(  10.0,  40.0)(  23.9,  27.6)(  38.0,  75.9)(  51.8,  58.1)
\psbezier(  51.8,  58.1)(  57.9,  50.5)(  63.4,  34.6)(  69.9,  25.2)
\psbezier(  69.9,  25.2)(  83.2,   6.2)(  97.3,  30.6)( 110.0,  25.0)
\psset{linewidth=0.5pt,linecolor=black}
\put(  8,40){\makebox(0,0)[r]{$s[0]$}}
\pscircle*( 10,40){0.8}
\psline( 10, 9)( 10,40)
\put( 29,60){\makebox(0,0){$s[1]$}}
\pscircle*( 30,50){0.8}
\psline( 30, 9)( 30,50)
\put( 50,68){\makebox(0,0){$s[2]$}}
\pscircle*( 50,60){0.8}
\psline( 50, 9)( 50,60)
\put( 72,35){\makebox(0,0){$s[3]$}}
\pscircle*( 70,25){0.8}
\psline( 70, 9)(70,25)
\put( 90,28){\makebox(0,0){$s[4]$}}
\pscircle*( 90,20){0.8}
\psline( 90, 9)( 90,20)
\put(110,32){\makebox(0,0){$s[5]$}}
\pscircle*(110,25){0.8}
\psline(110, 9)(110,25)
\psset{linewidth=1pt}
\psline{->}( 10,10)(118,10)
\psline{->}( 10,10)( 10,65)
\put( 122.0,  10.0){\makebox(0,0)[l]{$t$}}
\put(  10.0,  70.0){\makebox(0,0){$s(t)$}}
\put( 10, 7){\makebox(0,0)[t]{$0$}}
\put( 30, 7){\makebox(0,0)[t]{$T$}}
\put( 50, 7){\makebox(0,0)[t]{$2T$}}
\put( 70, 7){\makebox(0,0)[t]{$3T$}}
\put( 90, 7){\makebox(0,0)[t]{$4T$}}
\put(110, 7){\makebox(0,0)[t]{$5T$}}
\end{pspicture}

\caption{\label{fgGenrFunction}A signal $s(t)$ that is sampled with period $T$.}
\end{figure}
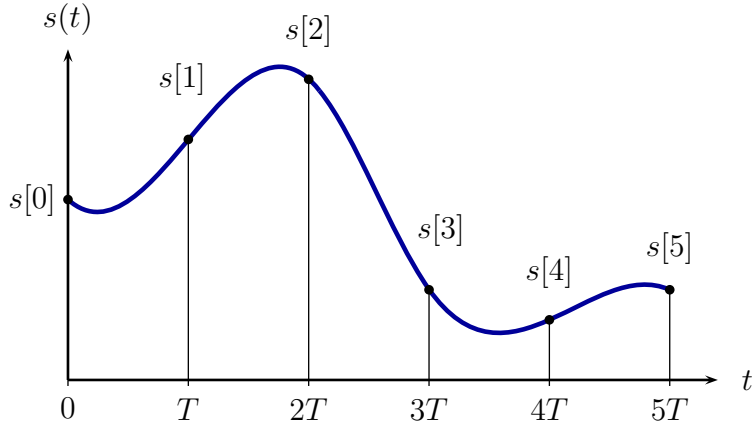

Even though this work is about two-dimensional images, to simplify the notation we assume that only separable filters
are used~\cite{Gonzalez92}, and thus consider only the one-dimensional case. In practice separable filters are almost
always used because of their much lower computational complexity.

In our analysis of sampling and reconstructing signals, we consider a function $s(t)$ with Fourier transform
\begin{equation}
 S(f) = \DIInt{s(t) \, \Cne{ft}}{t},
\end{equation}
that is sampled with period $T$, as shown in Figure~\ref{fgGenrFunction}, to generate the sequence
\begin{equation}
 s[n] = s(nT), \quad n \in \Integers.
\end{equation}
The Fourier transform of this sequence is
\begin{equation}
 \FT{S} = \DISum{n} s[n] \, e^{-j \omega n}.
\end{equation}
Note that we use the notation of reference~\cite{Oppenheim89}, using parenthesis for functions of real variables and
brackets for functions of integer variables.

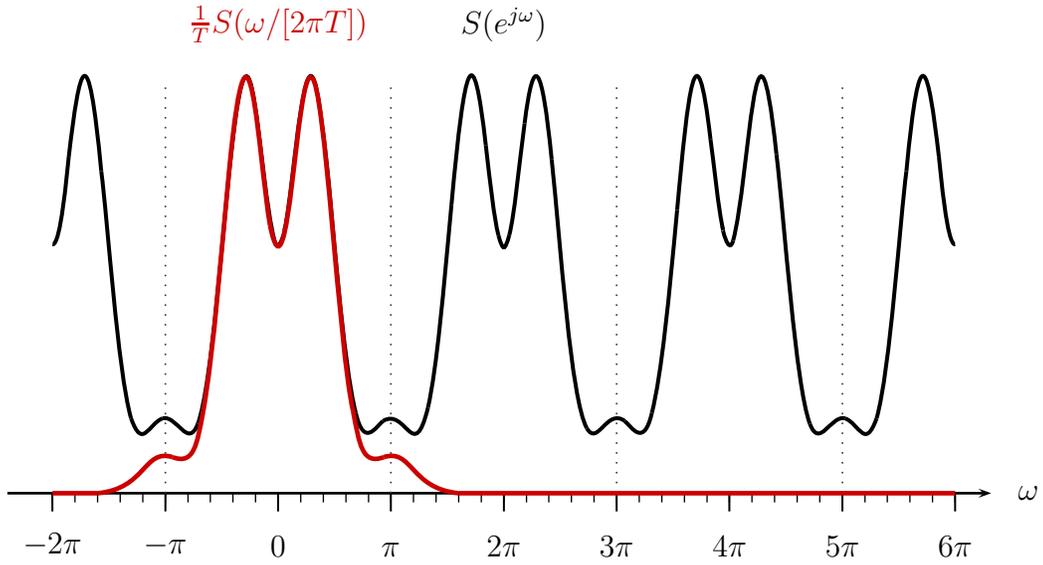
\begin{figure}[tb]
\centering
\setlength{\unitlength}{1.2mm}
\psset{unit=1.2mm}

\begin{pspicture}( 120.0,  70.0)
\psset{linewidth=0.75pt}
\psline[linewidth=1pt]{->}(5,10)(114,10)
\multiput(10,10)(12.5,0){ 9}{\psline(0,-2)}
\multiput(10,10)(  2.5,0){41}{\psline[linewidth=0.5pt](0,-1)}
\put( 10.00,3){\makebox(0,0)[b]{$-2\pi$}}
\put( 22.50,3){\makebox(0,0)[b]{$-\pi$}}
\put( 35.00,3){\makebox(0,0)[b]{$0$}}
\put( 47.50,3){\makebox(0,0)[b]{$\pi$}}
\put( 60.00,3){\makebox(0,0)[b]{$2\pi$}}
\put( 72.50,3){\makebox(0,0)[b]{$3\pi$}}
\put( 85.00,3){\makebox(0,0)[b]{$4\pi$}}
\put( 97.50,3){\makebox(0,0)[b]{$5\pi$}}
\put(110.00,3){\makebox(0,0)[b]{$6\pi$}}
\put( 117.0,  10.0){\makebox(0,0)[l]{$\omega$}}
\put(60,62){\makebox(0,0){$S(e^{j\omega})$}}
\put(35,62){\makebox(0,0){\color{asred} $\frac{1}{T} S(\omega/[2\pi T])$}}
\multiput(22.5,10)(25,0){4}{\psline[linewidth=0.75pt,linestyle=dotted,linecolor=darkgray](0,46)}
\psset{linewidth=1.5pt}
\psbezier(  10.0,  37.6)(  10.9,  37.4)(  11.6,  45.5)(  11.8,  47.3)
\psbezier(  11.8,  47.3)(  13.0,  57.5)(  13.9,  61.5)(  15.4,  45.7)
\psbezier(  15.4,  45.7)(  16.1,  40.4)(  17.2,  23.6)(  18.7,  18.3)
\psbezier(  18.7,  18.3)(  20.0,  14.7)(  20.6,  17.7)(  22.2,  18.3)
\psbezier(  22.2,  18.3)(  23.8,  18.7)(  25.0,  14.8)(  26.1,  17.7)
\psbezier(  26.1,  17.7)(  27.7,  22.0)(  28.9,  40.1)(  29.8,  47.6)
\psbezier(  29.8,  47.6)(  32.1,  69.0)(  32.8,  44.7)(  34.6,  38.0)
\psbezier(  34.6,  38.0)(  35.1,  36.2)(  35.9,  39.7)(  35.7,  39.2)
\psbezier(  35.7,  39.2)(  36.2,  41.1)(  36.9,  48.6)(  37.4,  51.7)
\psbezier(  37.4,  51.7)(  38.7,  60.6)(  39.4,  55.2)(  40.4,  46.4)
\psbezier(  40.4,  46.4)(  41.0,  41.0)(  42.2,  24.2)(  43.6,  18.9)
\psbezier(  43.6,  18.9)(  44.5,  15.0)(  45.8,  17.3)(  46.6,  18.0)
\psbezier(  46.6,  18.0)(  49.2,  19.6)(  50.0,  13.0)(  51.6,  19.5)
\psbezier(  51.6,  19.5)(  52.6,  24.0)(  53.3,  32.3)(  54.2,  42.0)
\psbezier(  54.2,  42.0)(  55.7,  58.4)(  56.6,  60.9)(  58.0,  48.6)
\psbezier(  58.0,  48.6)(  58.5,  44.3)(  58.9,  40.6)(  59.4,  38.6)
\psbezier(  59.4,  38.6)(  59.9,  36.8)(  60.1,  36.8)(  60.6,  38.6)
\psbezier(  60.6,  38.6)(  61.2,  41.0)(  61.6,  45.7)(  62.3,  51.0)
\psbezier(  62.3,  51.0)(  63.3,  59.0)(  64.2,  58.0)(  65.3,  47.0)
\psbezier(  65.3,  47.0)(  66.3,  37.5)(  67.4,  20.4)(  69.2,  17.2)
\psbezier(  69.2,  17.2)(  70.4,  15.1)(  71.4,  19.5)(  73.4,  18.0)
\psbezier(  73.4,  18.0)(  74.2,  17.3)(  75.5,  15.0)(  76.4,  18.9)
\psbezier(  76.4,  18.9)(  77.5,  23.3)(  78.3,  31.7)(  79.2,  41.4)
\psbezier(  79.2,  41.4)(  80.4,  54.5)(  81.3,  63.0)(  82.9,  49.2)
\psbezier(  82.9,  49.2)(  83.2,  47.2)(  83.8,  41.1)(  84.3,  39.2)
\psbezier(  84.3,  39.2)(  84.1,  39.7)(  84.9,  36.2)(  85.4,  38.0)
\psbezier(  85.4,  38.0)(  87.2,  44.7)(  87.9,  69.0)(  90.2,  47.6)
\psbezier(  90.2,  47.6)(  91.1,  40.1)(  92.3,  22.0)(  93.9,  17.7)
\psbezier(  93.9,  17.7)(  95.0,  14.8)(  96.2,  18.7)(  97.8,  18.3)
\psbezier(  97.8,  18.3)(  99.4,  17.7)( 100.0,  14.7)( 101.3,  18.3)
\psbezier( 101.3,  18.3)( 102.8,  23.6)( 103.9,  40.4)( 104.6,  45.7)
\psbezier( 104.6,  45.7)( 106.1,  61.5)( 107.0,  57.5)( 108.2,  47.3)
\psbezier( 108.2,  47.3)( 108.4,  45.5)( 109.1,  37.4)( 110.0,  37.6)
\psset{linewidth=1.75pt,linecolor=asred}
\psline(10,10)(15,10)
\psbezier(  15.0,  10.0)(  20.4,  10.5)(  20.1,  15.0)(  23.3,  14.0)
\psbezier(  23.3,  14.0)(  25.7,  13.1)(  26.0,  15.4)(  26.7,  19.1)
\psbezier(  26.7,  19.1)(  27.9,  25.9)(  28.8,  37.9)(  29.6,  46.1)
\psbezier(  29.6,  46.1)(  31.8,  68.6)(  32.8,  47.3)(  34.1,  39.7)
\psbezier(  34.1,  39.7)(  36.2,  29.3)(  37.0,  57.2)(  38.7,  56.1)
\psbezier(  38.7,  56.1)(  40.2,  55.5)(  41.7,  26.0)(  43.3,  19.1)
\psbezier(  43.3,  19.1)(  44.0,  15.4)(  44.3,  13.1)(  46.7,  14.0)
\psbezier(  46.7,  14.0)(  49.9,  15.0)(  49.6,  10.5)(  55.0,  10.0)
\psline(55,10)(110,10)
\end{pspicture}

\caption{\label{fgPerSpect}A signal spectrum and the Fourier Transform of its sampled sequence.}
\end{figure}

From the sampling theorem~\cite{Oppenheim89,Gonzalez92} we have, for $\omega=2 \pi fT$,
\begin{equation}
 \FT{S} = \Inv{T} \DISum{n} \Funct{S}{\frac{\omega}{2 \pi T} - \frac{n}{T}},
\end{equation}
or, equivalently,
\begin{equation}
 \FTf{S}{fT} = \Inv{T} \DISum{n} \Funct{S}{f - \frac{n}{T}}.
\end{equation}
Figure~\ref{fgPerSpect} shows in red an example of the bandlimited spectrum of a signal $s(t)$, and the corresponding
periodic spectrum of its sampled sequence. We can note that in this example the bandwidth of the signal is larger than
$1/(2T)$, so there is overlap in the shifted versions of $S(f)$, i.e., \textit{aliasing}~\cite{Oppenheim89}.

%

Given a reconstruction kernel $\phi(t)$ with Fourier transform $\Phi(f)$, we define a {\em reconstructed function}
\begin{equation}
 r(t) = \DISum{k} s[k] \, \Funct{\phi}{\frac{t}{T} - k},
\end{equation}
which has Fourier transform
\begin{eqnarray}
 R(f) & = & T \, \FTf{S}{fT} \, \Phi(fT) \\
  & = & \Phi(fT) \DISum{k} \Funct{S}{f - \frac{k}{T}}.
\end{eqnarray}

From this equation we can observe that if the signal $s(t)$ is strictly bandlimited, i.e., if its spectrum satisfies
\begin{equation}
 \label{eqBWLimt}
 S(f) = 0, \quad \forall f \not\in \Parth{-\Inv{2T},\Inv{2T}}
\end{equation}
then we can have $R(f)=S(f)$, and consequently $r(t)=s(t)$, if we use as $\phi(t)$ the box (ideal filter) kernel
\begin{equation}
 b(t) = \frac{\sin(\pi t)}{\pi t},
\end{equation}
which has Fourier transform
\begin{equation}
 B(f) = \left\{ \begin{array}{ll}
  1, & |f| < 1/2, \\
  1/2, & |f| = 1/2, \\
  0, & |f| > 1/2, \\
  \end{array} \right.
\end{equation}

If we resample $r(t)$ with period $\tau$, we have
\begin{equation}
 r[n] = r(n\tau) = \DISum{k} s[k] \, \Funct{\phi}{\frac{n\tau}{T} - k}
\end{equation}
and defining
\begin{equation}
 \gamma = \frac{T}{\tau},
\end{equation}
we obtain
\begin{equation}
 \FTf{R}{f} = \gamma \DISum{n} \Phi(\gamma[f-n]) \FTf{S}{\gamma[f-n]}.
\end{equation}

From this result it is clear that we can only have $\FT{R}=\FT{S}$ when $T=\tau$ and $\phi(t)=b(t)$. However, if $\tau\leq T$ and $s(t)$
satisfies~(\ref{eqBWLimt}), then we can preserve all information on $s(t)$ using $\phi(t)=b(t)$. If, on the
other hand, we have $\tau>T$, then spectrum of $R(f)$ has to satisfy
\begin{equation}
 \label{eqBWLimtT}
 R(f) = 0, \quad \forall f \not\in \Parth{-\Inv{2\tau},\Inv{2\tau}}
\end{equation}
to avoid aliasing. This can be done by using a time-stretched version $\phi(t) = \gamma b(\gamma t)$ as the reconstruction kernel.

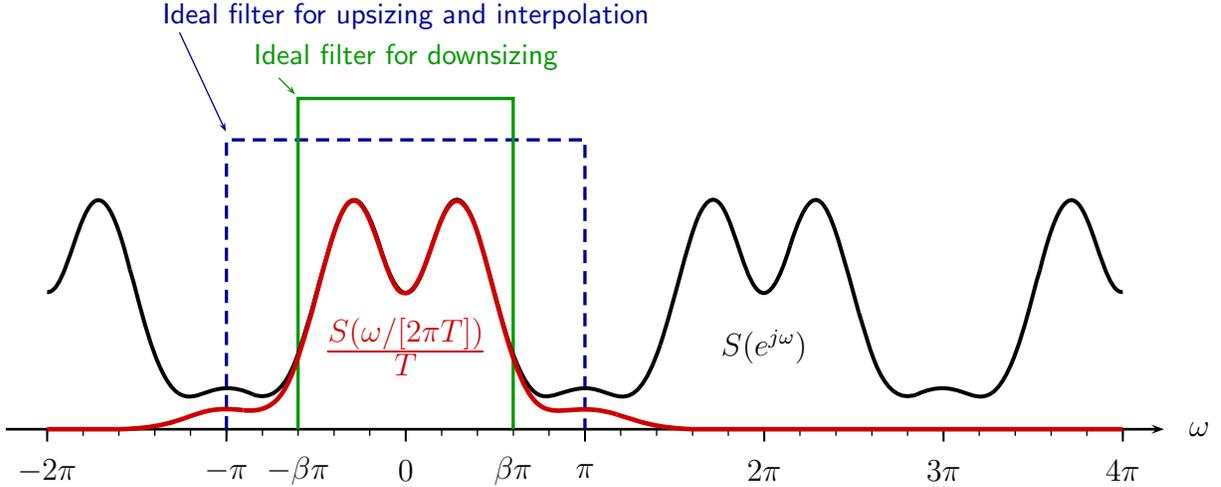
\begin{figure}[tb]
\centering
\setlength{\unitlength}{1.1mm}
\psset{unit=1.1mm}

\begin{pspicture}(150,70)
\psset{linewidth=0.75pt}
\psline[linewidth=1pt]{->}( 5, 10)(145,10)
\multiput( 10,10)(21.667,0){ 7}{\psline(0,-1.5)}
\multiput( 10,10)( 4.333,0){31}{\psline[linewidth=0.5pt](0,-0.8)}
\put( 10.00,5){\makebox(0,0){$-2\pi$}}
\put( 31.67,5){\makebox(0,0){$-\pi$}}
\put( 53.33,5){\makebox(0,0){$0$}}
\put( 75.00,5){\makebox(0,0){$\pi$}}
\put( 96.67,5){\makebox(0,0){$2\pi$}}
\put(118.33,5){\makebox(0,0){$3\pi$}}
\put(140.00,5){\makebox(0,0){$4\pi$}}
\put( 40.33,5){\makebox(0,0){$-\beta\pi$}}
\put( 66.33,5){\makebox(0,0){$\beta\pi$}}
\put(148,10){\makebox(0,0)[l]{$\omega$}}
\put(53.33,20){\makebox(0,0){\color{asred} $\frac{\displaystyle S(\omega/[2\pi T])}{\displaystyle T}$}}
\put(96.67,20){\makebox(0,0){$S(e^{j\omega})$}}
\psline[linewidth=0.5pt,linecolor=asblue]{->}(26,58)(31.6,46)
\psline[linewidth=1.25pt,linecolor=asblue,linestyle=dashed](31.67,10)(31.67,45)(75,45)(75,10)
\put(53.33,60){\makebox(0,0){\color{asblue} \small \sf Ideal filter for upsizing and interpolation}}
\psline[linewidth=0.5pt,linecolor=asgreen]{->}(38,52.5)(40,50.5)
\psline[linewidth=1.25pt,linecolor=asgreen](40.33,10)(40.33,50)(66.33,50)(66.33,10)
\put(53.33,55){\makebox(0,0){\color{asgreen} \small \sf Ideal filter for downsizing}}
\psset{linewidth=1.5pt}
\psbezier(  10.0,  26.6)(  13.4,  26.1)(  14.9,  49.9)(  20.0,  29.3)
\psbezier(  20.0,  29.3)(  21.0,  26.3)(  22.9,  16.4)(  25.8,  14.4)
\psbezier(  25.8,  14.4)(  28.3,  12.7)(  30.0,  16.5)(  34.9,  14.2)
\psbezier(  34.9,  14.2)(  39.1,  12.7)(  40.5,  19.8)(  41.9,  24.1)
\psbezier(  41.9,  24.1)(  45.3,  36.4)(  46.7,  43.2)(  50.2,  32.2)
\psbezier(  50.2,  32.2)(  55.7,  14.0)(  56.7,  45.9)(  61.3,  35.7)
\psbezier(  61.3,  35.7)(  62.7,  32.4)(  64.1,  25.7)(  65.9,  20.2)
\psbezier(  65.9,  20.2)(  69.1,  11.3)(  70.4,  14.4)(  74.3,  14.9)
\psbezier(  74.3,  14.9)(  78.4,  15.6)(  80.6,   9.9)(  84.1,  20.2)
\psbezier(  84.1,  20.2)(  85.9,  25.7)(  87.3,  32.4)(  88.7,  35.7)
\psbezier(  88.7,  35.7)(  93.3,  45.9)(  94.3,  14.0)(  99.8,  32.2)
\psbezier(  99.8,  32.2)( 103.3,  43.2)( 104.7,  36.4)( 108.1,  24.1)
\psbezier( 108.1,  24.1)( 109.5,  19.8)( 110.9,  12.7)( 115.1,  14.2)
\psbezier( 115.1,  14.2)( 120.0,  16.5)( 121.7,  12.7)( 124.2,  14.4)
\psbezier( 124.2,  14.4)( 127.1,  16.4)( 129.0,  26.3)( 130.0,  29.3)
\psbezier( 130.0,  29.3)( 135.1,  49.9)( 136.6,  26.1)( 140.0,  26.6)
\psset{linewidth=1.75pt,linecolor=asred}
\psline(10,10)(18.7,10)
\psbezier(  18.7,  10.0)(  27.7,  10.2)(  27.3,  13.1)(  33.7,  12.3)
\psbezier(  33.7,  12.3)(  38.3,  11.8)(  39.4,  15.4)(  41.5,  22.5)
\psbezier(  41.5,  22.5)(  43.0,  27.3)(  44.8,  36.3)(  46.6,  37.5)
\psbezier(  46.6,  37.5)(  49.2,  39.1)(  50.8,  26.8)(  53.0,  26.5)
\psbezier(  53.0,  26.5)(  56.4,  25.5)(  57.8,  45.1)(  61.8,  34.4)
\psbezier(  61.8,  34.4)(  62.8,  31.6)(  64.0,  26.5)(  65.0,  23.0)
\psbezier(  65.0,  23.0)(  69.4,   7.1)(  71.0,  14.3)(  78.8,  11.8)
\psbezier(  78.8,  11.8)(  82.0,  10.8)(  83.4,  10.2)(  88.0,  10.0)
\psline(88,10)(140,10)
\end{pspicture}

\caption{Box filters used for strictly bandlimited signals.}
\label{fgPerBoxFilter}
\end{figure}

Figure~\ref{fgPerBoxFilter} shows the two cases, when ideal (box) filters are used. When $\tau\leq T$ (upsampling
and interpolation) the filter response covers the whole original bandwidth, and when $\tau>T$ (downsampling) the
filter response is reduced by a factor $T/\tau$ to avoid aliasing.

In conclusion, when it is known that a signal that has been sampled with period $T$ needs to be resampled with
period $\tau$, we can use the kernels
\begin{equation}
 \phi(t) = \beta \, b(\beta t),
\end{equation}
where
\begin{equation}
 \label{eqBetaDefn}
 \beta = \min(1, T / \tau).
\end{equation}

%

\section{Kernels for Image Interpolation and Resizing}\label{ssInterp}

%
\begin{figure}[p]
\centering

\includegraphics[width=3in]{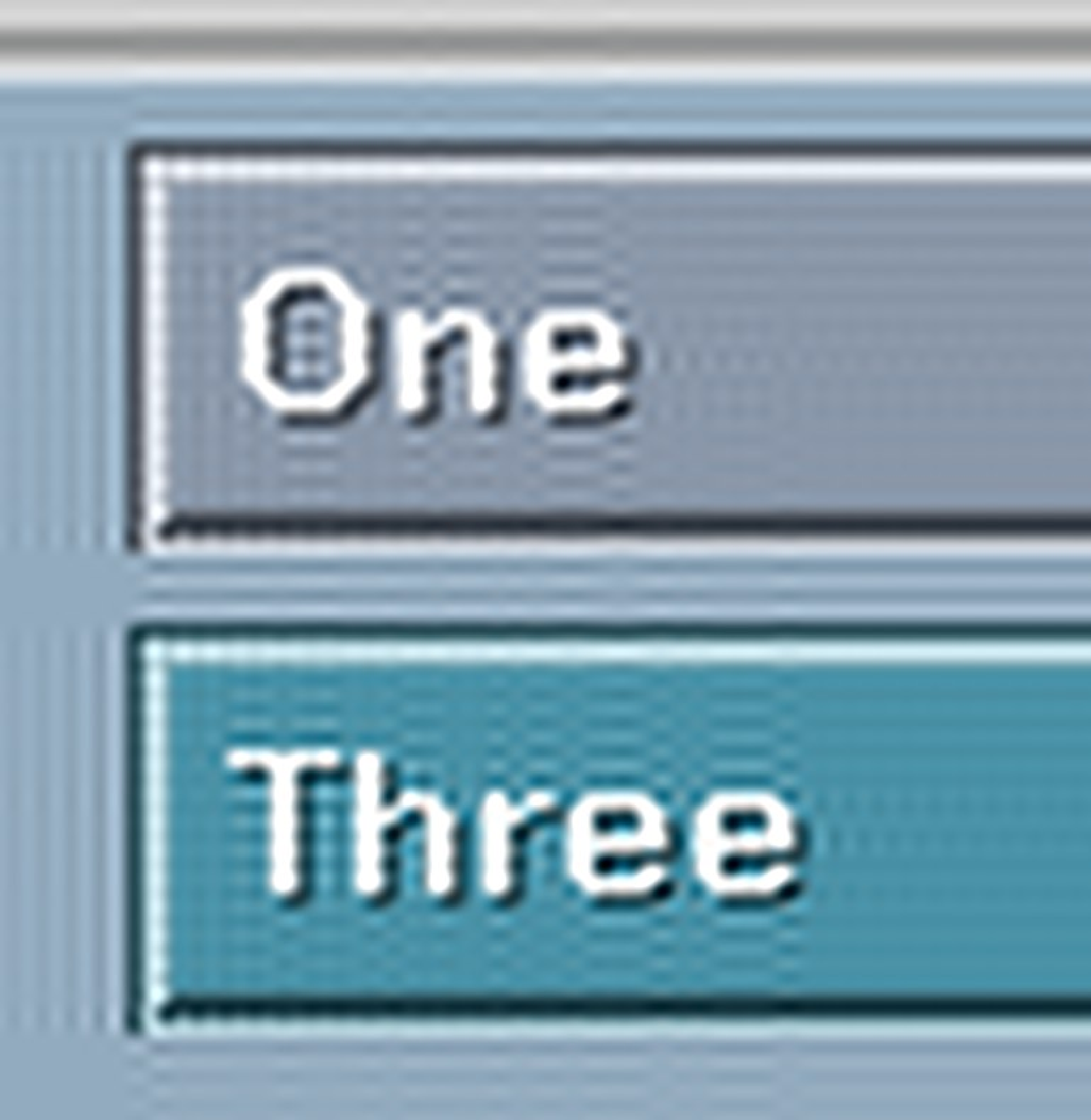}
\hspace{0.2in}
\includegraphics[width=3in]{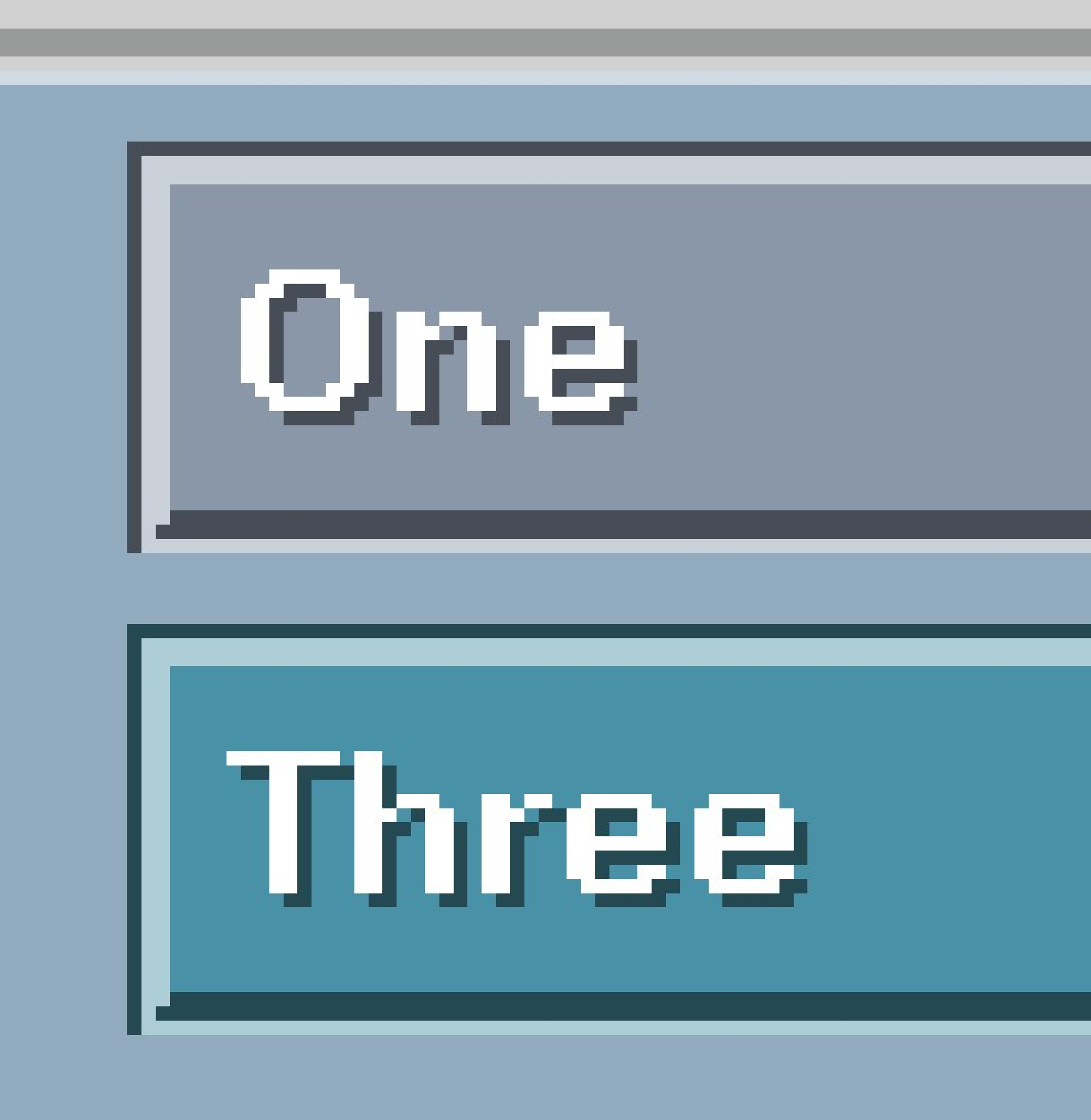}

(a) \hspace{3.1in} (b)

\vspace{0.2in}

\includegraphics[width=3in]{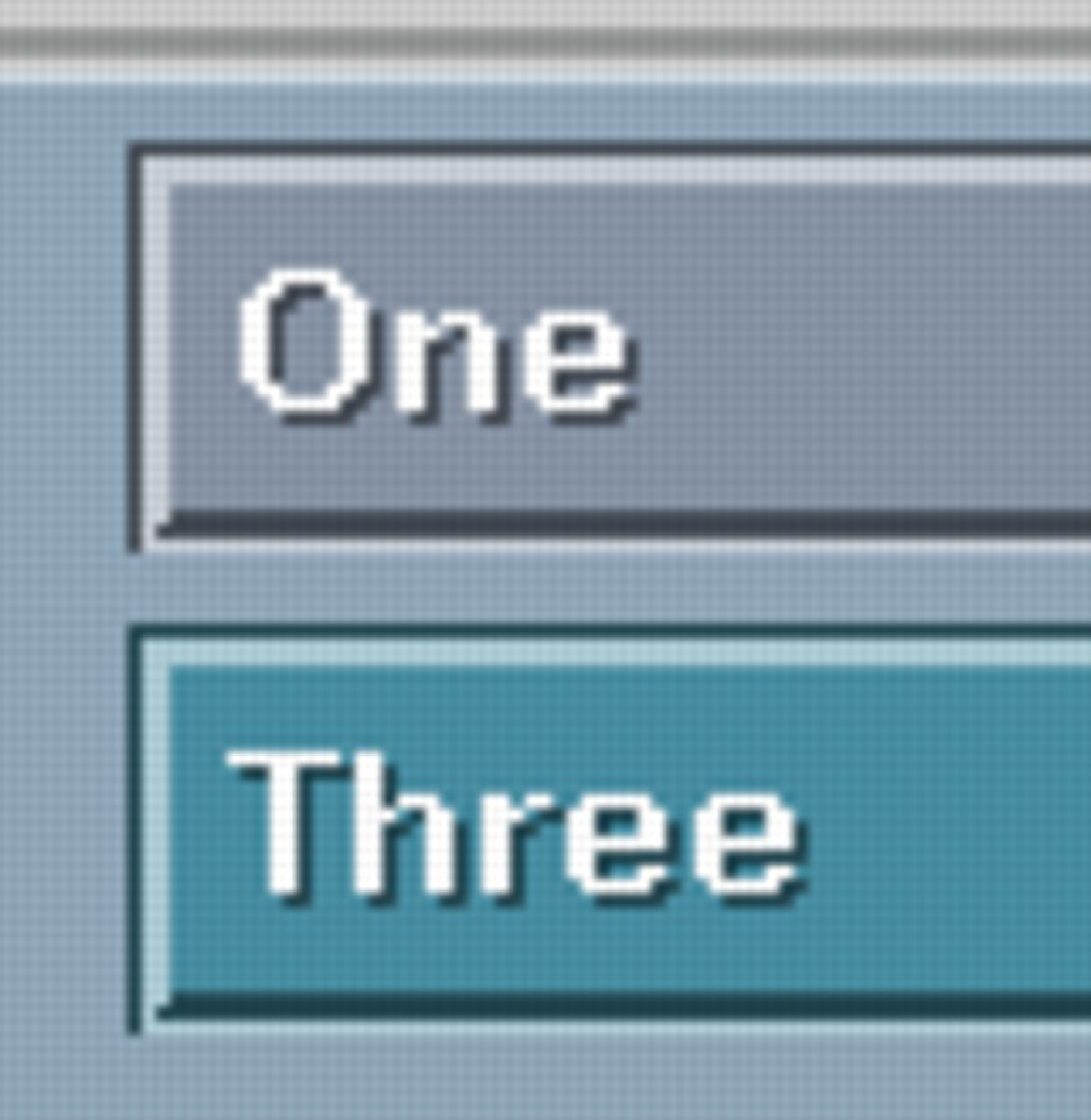}
\hspace{0.2in}
\includegraphics[width=3in]{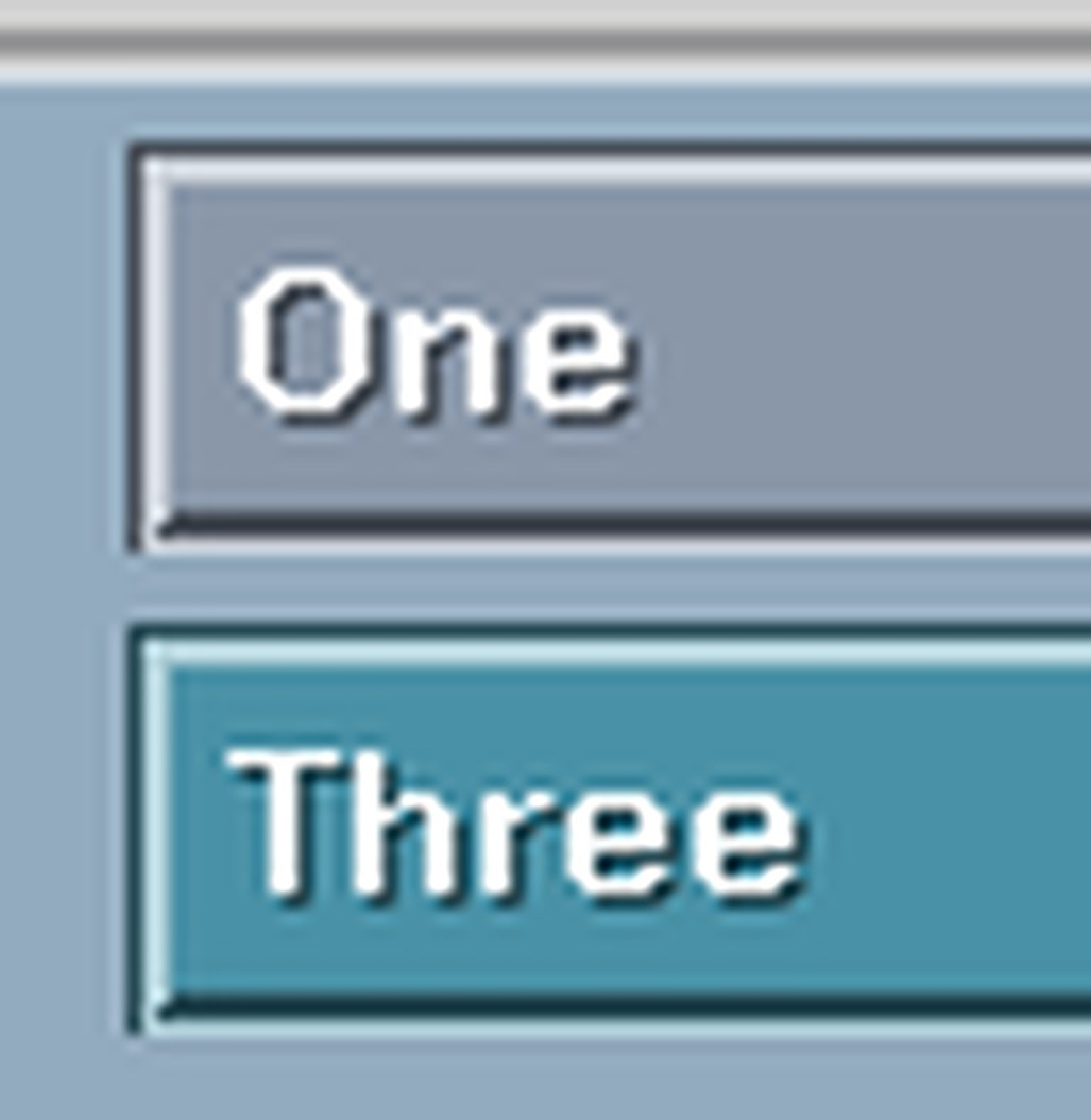}

(c) \hspace{3.1in} (d)

\caption{Examples of an image with text interpolated using different kernels. (a)~Ideal bandlimited filter; (b)~Pixel
replication kernel; (c)~``Screen door'' effect caused by insufficient attenuation of repeated versions of the spectrum;
(d)~cubic spline kernel.}

\label{fgIntpText}
\end{figure}

The theory presented in Section~\ref{ssSampling} was developed for analyzing exact recovery of sampled signals that
are strictly bandlimited, but in imaging applications we normally do not have signals that are truly bandlimited. One
common example is in images containing text, which define discontinuous signals. The application of ideal filters
to those images produces highly visible ``ringing'' artifacts, as shown in Figure~\ref{fgIntpText}(a). Thus, other
types of filters have been used for images~\cite{Harris78}--\cite{Meijering02}. For upsizing by integer factors,
the simplest technique is to just replicate the pixel values, but as can be seen in Figure~\ref{fgIntpText}(b),
the results are also visually quite bad.

Another problem occurs when the filters do not remove enough of the periodic versions in the spectrum of the sampled
signal. In this case we have the ``screen door'' effect, as shown in Figure~\ref{fgIntpText}(c). Much better results are
obtained using some kernels defined for general function interpolation. For instance, Figure~\ref{fgIntpText}(d) shows
the image obtained with cubic-spline interpolation.

%
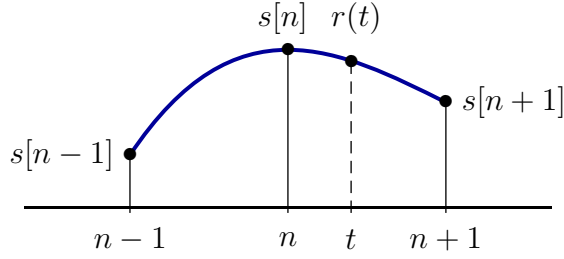
\begin{figure}[tb]
\centering
\setlength{\unitlength}{7mm}
\psset{unit=7mm}
\begin{pspicture}(10,5)(0,0.5)
\psset{linewidth=0.5pt}
\psline[linestyle=dashed](6.2,0.9)(6.2,3.77)
\psline(2,0.9)(2,2)
\psline(5,0.9)(5,4)
\psline(8,0.9)(8,3)
\thicklines
\put(0,1){\line(1,0){10}}
\psbezier[linecolor=asblue,linewidth=1.5pt](2,2)(4,5)(6,4)(8,3)
\put(6.2,3.77){\makebox(0,0){$\bullet$}}
\put(2,2){\makebox(0,0){$\bullet$}}
\put(5,4){\makebox(0,0){$\bullet$}}
\put(8,3){\makebox(0,0){$\bullet$}}
\put(6.2,0.4){\makebox(0,0){$t$}}
\put(2,0.4){\makebox(0,0){$n-1$}}
\put(5,0.4){\makebox(0,0){$n$}}
\put(8,0.4){\makebox(0,0){$n+1$}}
\put(6.3,4.6){\makebox(0,0)[c]{$r(t)$}}
\put(1.7,2){\makebox(0,0)[r]{$s[n-1]$}}
\put(4.9,4.6){\makebox(0,0)[c]{$s[n]$}}
\put(8.3,3){\makebox(0,0)[l]{$s[n+1]$}}
\end{pspicture}
\caption{Notation used for interpolation and resizing.}
\label{fgInterpDiag}
\end{figure}

Figure~\ref{fgInterpDiag} shows the basic notation we use. We assume $T=1$, and for a sequence of signal values $s[n]$,
and an interpolation kernel function $h(t)$,\footnote{In Section~\ref{ssSampling} we use $\phi(t)$ in the
analysis of strictly bandlimited signals. In the rest of this document we use $h(t)$ to identify the type of kernels
that are used in practice.}
the reconstructed value at point $t$ is defined as
\begin{equation}
 r(t) = \beta \, \DISum{i} s[i] h(\beta[t-i]).
\end{equation}
where $\beta$ is the normalized bandwidth of the signal, as defined by~(\ref{eqBetaDefn}). When downsampling the image
factor $\beta$ should be equal or smaller
than the reduction factor, which means that $0<\beta<1$. When the image is upsampled or rotated, we have $\beta=1$.

This series can be also defined as
\begin{equation}
 r(t) = \beta \, \sum_{k \in S_h} s[k+\Floor{t}] h(\beta[t-k-\Floor{t}]),
\end{equation}
where $S_h$ is a range of integers defined by the condition $h(\beta t)\neq 0$. In practice, we can use the
condition $|h(\beta t)|<\epsilon$, where $\epsilon$ is sufficiently small.

Using the new notation with interpolation kernels, if we resample $r(t)$ with period $\tau$, we have the convolution
\begin{equation}
 r[n] = r(n\tau) = \beta \, \InfSum{k=\infty} s[\eta_n - k] h_{\alpha_n,\beta}[k],
\end{equation}
where
\begin{eqnarray}
 \eta_n & = & \Floor{n\tau}, \\
 \alpha_n  & = & n\tau - \Floor{n\tau}, \\
 h_{\alpha,\beta}[k] & = & h(\beta[k + \alpha]).
\end{eqnarray}

This means that the new sequence has each sample defined by a convolution sum, but the coefficients can be different
in each case. When the ratio $T/\tau$ is integer or rational, the sets of coefficients occur periodically, and
we have a multi-rate system.

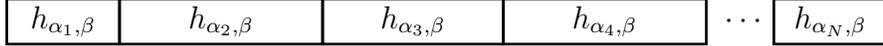
\begin{figure}
\setlength{\unitlength}{0.75mm}
\psset{unit=0.75mm}
\centering
\begin{pspicture}(160,20)
\psset{linewidth=1pt,dimen=middle}
\put(  0, 0){\psframe(20, 8) \makebox(20, 8){$h_{\alpha_1,\beta}$}}
\put( 20, 0){\psframe(36, 8) \makebox(36, 8){$h_{\alpha_2,\beta}$}}
\put( 56, 0){\psframe(32, 8) \makebox(32, 8){$h_{\alpha_3,\beta}$}}
\put( 88, 0){\psframe(36, 8) \makebox(36, 8){$h_{\alpha_4,\beta}$}}
\put(136, 0){\psframe(20, 8) \makebox(20, 8){$h_{\alpha_N,\beta}$}}
\put(131, 4){\makebox(0,0){$\cdots$}}
\end{pspicture}

\caption{Arrays with sequence of filter coefficients that are pre-computed for image resizing, and repeatedly used in the vertical
(or horizontal) resampling computations.}
\label{fgFilterCoeffArray}
\end{figure}

For image applications it is commonly more efficient to pre-compute and store sequentially all the coefficients of these
convolutions in a single array (cf. Figure~\ref{fgFilterCoeffArray}), which are later repeatedly used when the images lines
are resampled. Only one set is needed for the vertical resizing, and another for the horizontal resizing.

%

\section{New Kernel for Interpolation and Resizing}\label{ssNewKernel}

The family of functions that we propose for interpolation and resizing has only two parameters, $\chi$ and $\eta$, and
is defined by
\begin{equation}
 h_s(t; \chi,\eta) = \Sinc{t} \, 
   \Funct{\cosh}{\frac{\sqrt{2 \eta} \pi \chi t}{2-\eta}}
   e^{-[\pi \chi t /(2-\eta)]^2}
\end{equation}
where
\begin{equation}
 \Sinc{t} = \left\{ \begin{array}{ll}
  1, & t = 0, \\
  \sin(\pi t)/(\pi t), & t \neq 0, \\
 \end{array} \right.
\end{equation}

The corresponding Fourier transforms are
\begin{equation}
 H_s(f; \chi,\eta) = \Funct{P_s}{\frac{[2f+1][2-\eta]}{\sqrt{2} \chi}; \eta} -
    \Funct{P_s}{\frac{[2f-1][2-\eta]}{\sqrt{2} \chi}; \eta}
\end{equation}
where
\begin{equation}
 \label{eqReComplxErfc}
 P_s(f; \eta) = \frac{e^{\eta/2}}{\sqrt{2 \pi}} \int_0^f e^{-\phi^2/2} \cos(\sqrt{\eta}\phi) \, \mathrm{d}\phi
\end{equation}

%
\begin{figure}[tb]
\centering
\setlength{\unitlength}{1.2mm}
\psset{unit=1.2mm}

\input{fg_ir_02.tex}

\input{fg_fr_02.tex}

\caption{Impulse and frequency responses for some selected parameters, $\chi=0.2$.}
\label{fgRespGraphsA}
\end{figure}

%
\begin{figure}[tb]
\centering
\setlength{\unitlength}{1.2mm}
\psset{unit=1.2mm}

\input{fg_ir_03.tex}

\input{fg_fr_03.tex}

\caption{Impulse and frequency responses for some selected parameters, $\chi=0.3$.}
\label{fgRespGraphsB}
\end{figure}

%
\begin{figure}[tb]
\centering
\setlength{\unitlength}{1.2mm}
\psset{unit=1.2mm}

\input{fg_ir_04.tex}

\input{fg_fr_04.tex}

\caption{Impulse and frequency responses for some selected parameters, $\chi=0.4$.}
\label{fgRespGraphsC}
\end{figure}

Figures \ref{fgRespGraphsA}, \ref{fgRespGraphsB}, and \ref{fgRespGraphsC} show some examples. Note that all graphs have $|h_s(t; \chi,\eta)|$
and $|H_s(f; \chi,\eta)|$ in decibels. We can observe that $\chi$ basically controls the width of the transition band,
and $\eta$ affects the height of the first sidelobe. Thus, when $\chi\rightarrow 0$ the functions $H_s(f; \chi,\eta)$
converge to the ideal lowpass filter, i.e.,
\begin{equation}
 \lim_{\chi\rightarrow 0} H_s(f; \chi,\eta) = \begin{cases}
  1, & |f| < 1/2, \\
  1/2, & |f| = 1/2, \\
  0, & |f| > 1/2.
 \end{cases}
\end{equation}

While there is no closed-form expression for integral~(\ref{eqReComplxErfc}) (related to the complex-valued error
function), we used the time-frequency properties of the functions, and found that it can approximated with absolute
error smaller than $10^{-16}$ using
\begin{equation}
 P_s(f; \eta) \approx \begin{cases}
   1/2, & f > 8, \\
   \displaystyle \frac{f}{17} + \sum_{n=1}^{22} \frac{e^{-\omega^2 n^2}}{\pi n} \cosh(2 \omega n \sqrt{\eta})
     \sin(2 \omega n f), & |f| \leq 8, \\
  -1/2, & f < -8,
  \end{cases}
\end{equation}
where $\omega=\pi/17$.

\section{Desirable Features}\label{ssFeatures}

There are some important features---not all simultaneously achievable---that are desirable for the kernel functions
used for creating the discrete-time filters.

\subsection{Flexibility}

It is necessary to recognize that different types of images (natural, medical, synthetic, etc.) have different
requirements. While we have a variety of theoretical tools developed for the analysis and design of interpolation and
resizing filters, in most cases it is still essential to experiment several filters, and visually inspect the results.

The proposed kernels are meant to allow imaging professionals to try different filters more easily. In fact, to make
their performance and visual quality easier to predict, they can closely approximate other commonly used kernels. This
way, it is possible to start with those approximations, and see how the image quality is altered after changes
in the parameters.

Table~\ref{tbParApprox} shows some sets of parameters that can be used for these approximations, and Figures
\ref{fgApproxGraphsA} and \ref{fgApproxGraphsB} show how good the approximation are. (More details in
Section~\ref{scResults}.)

\clearpage

\subsection{Intuitive Controls}

While experimenting, it is desirable to be able to finely tune the filter's response. Some kernels provide very little
control, being defined only for some discrete parameters, like ``order.'' Others are defined by parameters related to
approximation theory, which may have limited relation to the properties of natural images.

We defined the parameters of our kernel in a way that makes its equations somewhat more complicated, but aiming to make
them much more intuitive. The parameter $\chi$ is defined to be the main control for achieving a compromise between
blurring, aliasing and ringing artifacts. If it it too small, we have nearly ideal filters, which create ringing
artifacts around edges. If it is too large we some have blurring and the screen door effect. The amount of aliasing depends
also on the parameter $\eta$, which controls the height of the sidelobe. These properties can be seen in Figures
\ref{fgRespGraphsA} to \ref{fgRespGraphsC}.

\subsection{Symmetry and Exact Interpolation}

For imaging applications it is necessary to use linear phase filters, and commonly interpolation functions have even
symmetry, i.e., $h(t)=h(-t)$. Under the assumption that pixel values corresponds to samples of a strictly bandlimited
signal, we would like to not change the values that are already know, and this is achieved when
\begin{equation}
 h(0) = 1, \quad h(n) = 0, \quad n = \pm 1, \pm 2, \ldots
\end{equation}
Image signals are certainly not strictly bandlimited, but the property is still useful because it implies that for all
$f$ we have
\begin{equation}
 \Funct{H_d}{e^{j2\pi f}} = \sum_{n=-\infty}^{\infty} H(f-n) = 1,
\end{equation}
i.e., we know that the gain for signal plus aliasing always adds to one. Our kernels satisfy this property because
they belong to the class of functions created by multiplying sinc($t$) with another function.

\subsection{Good Response with Small Spatial Support}

When considering the filtering computational complexity, it is good to use filters with small numbers of taps. Since in
imaging applications we need to avoid ringing resulting from lowpass filters with steep transition, good interpolation
results had been obtained with very short filters.

Our kernels were chosen such that $h_s(t; \chi,\eta)$ and $H_s(f; \chi,\eta)$ have very fast asymptotic decay. In fact,
there are constants $a,b,c$, and $d$ such that
\begin{eqnarray*}
 |h_s(t; \chi,\eta)| & \leq & a \, e^{-b t^2} \\
 |H_s(f; \chi,\eta)| & \leq & c \, e^{-d f^2}.
\end{eqnarray*}
Thus, while $h_s(t; \chi,\eta)$ strictly has infinite support, the very fast decay makes it
easy to find where to truncate the response without significantly changing the filter's performance (cf. Figures
\ref{fgRespGraphsA} to \ref{fgRespGraphsC}). This approach tends to yield somewhat longer filter responses, but it is
more convenient for obtaining downsampling filters, which need to be more carefully designed.

\subsection{Good Performance For Both Interpolation and Downsampling}\label{ssInterpDown}

One of the most natural requirements in interpolation and resizing is that when applied to an image with a constant
pixel value, it should always create another image with the same value. Thus, a kernel's DC response, i.e., the function
obtained when $s[n]\equiv1$ should be also identical to one. This is possible only when we have an exact {\em partition
of unity}:
\begin{equation}
 \label{eq:UnityPartT}
 \sum_{k=-\infty}^{\infty} \beta \cdot h(\beta \cdot [t-k]) = 1, \quad t \in [0,1).
\end{equation}
In the frequency domain this corresponds to
\begin{equation}
 \label{eq:UnityPartF}
 H(0) + 2 \sum_{n=1}^{\infty} \Funct{H}{\frac{n}{\beta}} \, \cos(2 \pi n t) = 1, \quad t \in [0,1).
\end{equation}

Interpolation kernels commonly satisfy this property exactly by having $H(0)=1$, and $H(n)=0$, $n=1,2,\ldots$ Others
provide very good approximation with very small values of $|H(0)-1|$ and $|H(n)|$, $n=1,2,\ldots$

The problem of using interpolation functions for downsampling is that while the condition above may be satisfied exactly
for $\beta=1$, it may not be a good approximation when $\beta<1$.

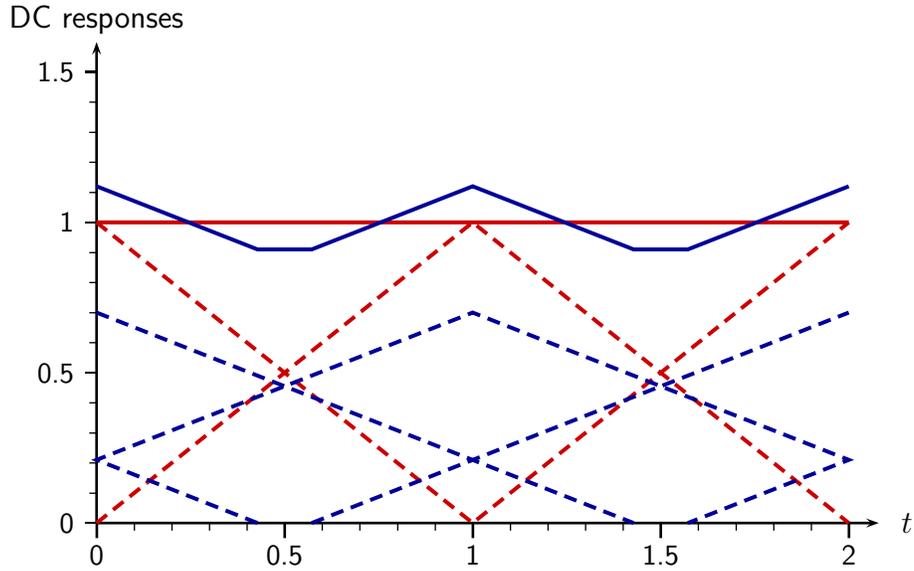
\begin{figure}[p]
\centering
\setlength{\unitlength}{1mm}
\psset{unit=1mm}

\setlength{\unitlength}{1mm}

\begin{pspicture}(120,80)
\psset{linewidth=1pt}
\psline{->}(10,10)(114,10)
\multiput(10,10)(25,0){5}{\psline(0,-2)}
\multiput(10,10)(5,0){21}{\psline[linewidth=0.5pt](0,-1)}
\put( 10,7){\makebox(0,0)[t]{\small \sf 0}}
\put( 35,7){\makebox(0,0)[t]{\small \sf 0.5}}
\put( 60,7){\makebox(0,0)[t]{\small \sf 1}}
\put( 85,7){\makebox(0,0)[t]{\small \sf 1.5}}
\put(110,7){\makebox(0,0)[t]{\small \sf 2}}
\put(117,10){\makebox(0,0)[l]{$t$}}
\psline{->}(10,10)(10,74)
\multiput(10,10)(0,20){4}{\line(-1,0){1.5}}
\multiput(10,10)(0,4){16}{\psline[linewidth=0.5pt](-1,0)}
\put(7,10){\makebox(0,0)[r]{\small \sf 0}}
\put(7,30){\makebox(0,0)[r]{\small \sf 0.5}}
\put(7,50){\makebox(0,0)[r]{\small \sf 1}}
\put(7,70){\makebox(0,0)[r]{\small \sf 1.5}}
\put(10,77){\makebox(0,0){\sf DC responses}}
\psset{linewidth=1.5pt,linecolor=asred,linestyle=dashed}
\psline[linestyle=solid](10,50)(110,50)
\psline(10,50)(60,10)(110,50)
\psline(10,10)(60,50)(110,10)
\psset{linecolor=asblue}
\psline[linestyle=solid](10,54.8)(31.4,46.4)(38.6,46.4)(60,54.8)(81.4,46.4)(88.6,46.4)(110,54.8)
\psline(10,38)(81.4,10)
\psline(38.6,10)(110,38)
\psline(31.4,10)(10,18.4)(60,38)(110,18.4)(88.6,10)
\end{pspicture}

\caption{Sum of linear interpolation (tent function) kernels when $s[n]\equiv1$. Dashed red lines correspond to shifted
components and continuous red line to their sum when $\beta=1$. The blue lines correspond to case $\beta=0.7$.}
\label{fgLinIntpSum}
\end{figure}

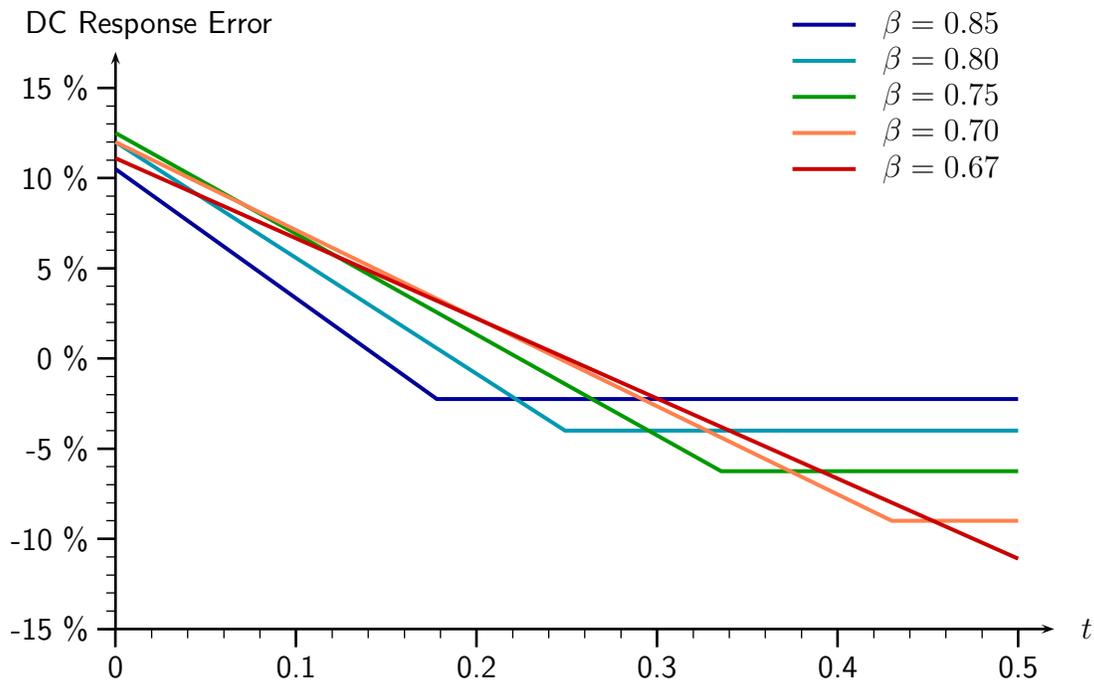
\begin{figure}[p]
\centering
\setlength{\unitlength}{1.2mm}
\psset{unit=1.2mm}

\begin{pspicture}(120,80)
\psset{linewidth=1pt}
\multiput(10,10)(4,0){26}{\psline[linewidth=0.5pt](0,-1)}
\multiput(10,10)(0,2){31}{\psline[linewidth=0.5pt](-1,0)}
\psline{->}(10,10)(114,10)
\multiput(10,10)(20,0){6}{\psline(0,-2)}
\put( 10,7){\makebox(0,0)[t]{\sf 0}}
\put( 30,7){\makebox(0,0)[t]{\sf 0.1}}
\put( 50,7){\makebox(0,0)[t]{\sf 0.2}}
\put( 70,7){\makebox(0,0)[t]{\sf 0.3}}
\put( 90,7){\makebox(0,0)[t]{\sf 0.4}}
\put(110,7){\makebox(0,0)[t]{\sf 0.5}}
\put(117,10){\makebox(0,0)[l]{$t$}}
\psline{->}(10,10)(10,74)
\multiput(10,10)(0,10){7}{\psline(-2,0)}
\put(7,10){\makebox(0,0)[r]{\sf -15 \%}}
\put(7,20){\makebox(0,0)[r]{\sf -10 \%}}
\put(7,30){\makebox(0,0)[r]{\sf  -5 \%}}
\put(7,40){\makebox(0,0)[r]{\sf   0 \%}}
\put(7,50){\makebox(0,0)[r]{\sf   5 \%}}
\put(7,60){\makebox(0,0)[r]{\sf  10 \%}}
\put(7,70){\makebox(0,0)[r]{\sf  15 \%}}
\put(0,77){\makebox(0,0)[l]{\sf DC Response Error}}
\psset{linewidth=1.5pt}
\psline[linecolor=asblue](85,77)(92,77)
\put(95,77){\makebox(0,0)[l]{$\beta=0.85$}}
\psline[linecolor=ascyan](85,73)(92,73)
\put(95,73){\makebox(0,0)[l]{$\beta=0.80$}}
\psline[linecolor=asgreen](85,69)(92,69)
\put(95,69){\makebox(0,0)[l]{$\beta=0.75$}}
\psline[linecolor=asorange](85,65)(92,65)
\put(95,65){\makebox(0,0)[l]{$\beta=0.70$}}
\psline[linecolor=asred](85,61)(92,61)
\put(95,61){\makebox(0,0)[l]{$\beta=0.67$}}
\psline[linecolor=asblue](10,61)(45.6,35.5)(110,35.5)
\psline[linecolor=ascyan](10,64)(59.8,32.0)(110,32)
\psline[linecolor=asgreen](10,65)(77.1,27.5)(110,27.5)
\psline[linecolor=asorange](10,64)(96.0,22.0)(110,22.0)
\psline[linecolor=asred](10,62.2)(110,17.8)
\end{pspicture}

\caption{Error in the DC response for linear interpolation kernels.}
\label{fgLinIntpPI}
\end{figure}

For example, let us consider the kernel for linear interpolation (tent function)
\begin{equation}
 h_l(t) = \begin{cases}
  1 - |t|, & |t| < 1, \\
  0, & |t| \geq 1,
 \end{cases}
\end{equation}
Its DC responses for $\beta=1$ and $\beta=0.7$ are shown in Figure~\ref{fgLinIntpSum}. We can observe that this kernel
satisfies the partition of unity condition when $\beta=1$, but is clearly inadequate when $\beta=0.7$. In fact, in
the latter case, if the resampling offset is zero, then the sampled response will alternate between values that are
about 10\% too large, and 10\% too small, i.e., an oscillation of about 20\%! Figure~\ref{fgLinIntpPI} shows how the
DC response error varies for other values of $\beta$.

%

%

Here we see one of the main advantages of kernel $h_s(t; \chi,\eta)$. With the proper choice of $\chi$ and $\eta$,
the very fast decay guarantees that $|H_s(1/b; \chi,\eta)|$ is very small for $\beta<1$, which means that it is also
good as a downsampling filter. In the next section we present some example of this property and comparisons of the DC
response errors (e.g., Figures \ref{fgSplineDCError} and \ref{fgLanczosDCError}).

\section{Approximation of Other Kernels}\label{scResults}

\begin{figure}[tbp]
\centering
\setlength{\unitlength}{1.2mm}
\psset{unit=1.2mm}

\begin{pspicture}(120,75)
\psset{linewidth=1pt}
\psline{->}(  10.0,  10.0)( 114.0,  10.0)
\multiput(  10.0,  10.0)(  16.667,   0.000){ 7}{\psline(   0.0,  -2.0)}
\multiput(  10.0,  10.0)(   3.333,   0.000){31}{\psline[linewidth=0.5pt](   0.0,  -1.0)}
\put(  10.0,   6.5){\makebox(0,0)[t]{\small \sf 0.00}}
\put(  26.7,   6.5){\makebox(0,0)[t]{\small \sf 0.25}}
\put(  43.3,   6.5){\makebox(0,0)[t]{\small \sf 0.50}}
\put(  60.0,   6.5){\makebox(0,0)[t]{\small \sf 0.75}}
\put(  76.7,   6.5){\makebox(0,0)[t]{\small \sf 1.00}}
\put(  93.3,   6.5){\makebox(0,0)[t]{\small \sf 1.25}}
\put( 110.0,   6.5){\makebox(0,0)[t]{\small \sf 1.50}}
\put( 117.0,  10.0){\makebox(0,0)[l]{\sf $f$}}
\psline{->}(  10.0,  10.0)(  10.0,  64.0)
\multiput(  10.0,  10.0)(   0.000,  12.500){ 5}{\psline(  -2.0,   0.0)}
\multiput(  10.0,  10.0)(   0.000,   3.125){17}{\psline[linewidth=0.5pt](  -1.0,   0.0)}
\put(   6.5,  10.0){\makebox(0,0)[r]{\small \sf -80 dB}}
\put(   6.5,  22.5){\makebox(0,0)[r]{\small \sf -60 dB}}
\put(   6.5,  35.0){\makebox(0,0)[r]{\small \sf -40 dB}}
\put(   6.5,  47.5){\makebox(0,0)[r]{\small \sf -20 dB}}
\put(   6.5,  60.0){\makebox(0,0)[r]{\small \sf   0 dB}}
\put(  10.0,  67.0){\makebox(0,0){\sf $|H(f)|$}}
\psset{linewidth=1.75pt,linecolor=asorange}
\psbezier(  65.5,  10.0)(  65.5,  11.9)(  65.5,  12.8)(  65.5,  15.0)
\psbezier(  65.5,  15.0)(  65.4,  16.7)(  65.4,  18.2)(  65.4,  20.0)
\psbezier(  65.4,  20.0)(  65.3,  21.9)(  65.3,  23.4)(  65.2,  24.9)
\psbezier(  65.2,  24.9)(  65.1,  26.7)(  65.0,  28.2)(  64.8,  29.9)
\psbezier(  64.8,  29.9)(  64.6,  30.9)(  64.4,  32.7)(  63.9,  34.8)
\psbezier(  63.9,  34.8)(  63.6,  35.7)(  63.1,  37.5)(  62.2,  39.5)
\psbezier(  62.2,  39.5)(  61.4,  41.4)(  60.4,  43.0)(  59.8,  43.8)
\psbezier(  59.8,  43.8)(  58.8,  45.2)(  57.9,  46.4)(  56.7,  47.6)
\psbezier(  56.7,  47.6)(  55.2,  49.2)(  53.7,  50.3)(  52.9,  50.9)
\psbezier(  52.9,  50.9)(  51.2,  52.2)(  49.7,  53.2)(  48.8,  53.7)
\psbezier(  48.8,  53.7)(  46.9,  54.7)(  45.2,  55.5)(  44.3,  55.8)
\psbezier(  44.3,  55.8)(  42.3,  56.7)(  40.6,  57.2)(  39.6,  57.5)
\psbezier(  39.6,  57.5)(  37.6,  58.1)(  35.8,  58.5)(  34.8,  58.7)
\psbezier(  34.8,  58.7)(  32.7,  59.1)(  30.9,  59.3)(  29.9,  59.4)
\psbezier(  29.9,  59.4)(  27.8,  59.7)(  25.9,  59.8)(  24.9,  59.8)
\psbezier(  24.9,  59.8)(  23.2,  59.9)(  21.7,  60.0)(  20.0,  60.0)
\psbezier(  20.0,  60.0)(  18.4,  60.0)(  17.0,  60.0)(  15.0,  60.0)
\psbezier(  15.0,  60.0)(  13.4,  60.1)(  12.0,  60.1)(  10.0,  60.1)
\psbezier(  70.7,  35.5)(  69.4,  35.5)(  68.2,  34.7)(  67.3,  32.8)
\psbezier(  67.3,  32.8)(  66.6,  31.3)(  66.3,  29.4)(  66.2,  28.4)
\psbezier(  66.2,  28.4)(  66.0,  26.7)(  65.8,  24.8)(  65.8,  23.8)
\psbezier(  65.8,  23.8)(  65.7,  22.4)(  65.7,  21.0)(  65.6,  19.2)
\psbezier(  65.6,  19.2)(  65.6,  17.8)(  65.6,  16.4)(  65.6,  14.6)
\psbezier(  65.6,  14.6)(  65.5,  12.9)(  65.6,  11.0)(  65.5,  10.0)
\psbezier(  79.9,  10.0)(  79.9,  11.0)(  79.9,  12.8)(  79.8,  14.9)
\psbezier(  79.8,  14.9)(  79.7,  16.8)(  79.6,  18.2)(  79.5,  19.7)
\psbezier(  79.5,  19.7)(  79.4,  21.4)(  79.2,  22.9)(  78.9,  24.5)
\psbezier(  78.9,  24.5)(  78.5,  26.2)(  78.2,  27.6)(  77.5,  29.2)
\psbezier(  77.5,  29.2)(  77.1,  30.1)(  76.4,  31.7)(  74.9,  33.3)
\psbezier(  74.9,  33.3)(  74.3,  34.0)(  72.6,  35.5)(  70.7,  35.5)
\psbezier(  95.6,  10.0)(  95.4,  12.1)(  95.6,  11.0)(  95.3,  14.7)
\psbezier(  95.3,  14.7)(  95.2,  16.3)(  95.0,  17.7)(  94.7,  19.3)
\psbezier(  94.7,  19.3)(  94.4,  20.9)(  94.1,  22.3)(  93.5,  23.8)
\psbezier(  93.5,  23.8)(  93.1,  24.6)(  92.4,  26.3)(  91.0,  27.8)
\psbezier(  91.0,  27.8)(  90.4,  28.4)(  89.1,  29.7)(  87.1,  30.0)
\psbezier(  87.1,  30.0)(  85.6,  30.2)(  84.1,  29.7)(  83.0,  28.2)
\psbezier(  83.0,  28.2)(  81.9,  26.9)(  81.4,  24.9)(  81.1,  24.0)
\psbezier(  81.1,  24.0)(  80.8,  22.5)(  80.6,  21.2)(  80.5,  19.3)
\psbezier(  80.5,  19.3)(  80.3,  17.7)(  80.2,  15.7)(  80.2,  14.7)
\psbezier(  80.2,  14.7)(  80.1,  13.8)(  80.1,  10.1)(  80.1,  10.0)
\psbezier( 110.0,  19.2)( 109.4,  20.9)( 108.8,  22.2)( 107.6,  23.6)
\psbezier( 107.6,  23.6)( 107.0,  24.3)( 105.6,  25.8)( 103.5,  26.1)
\psbezier( 103.5,  26.1)( 101.5,  26.3)( 100.0,  25.5)(  99.1,  24.3)
\psbezier(  99.1,  24.3)(  97.9,  23.0)(  97.3,  20.9)(  97.0,  19.9)
\psbezier(  97.0,  19.9)(  96.6,  18.1)(  96.3,  16.0)(  96.2,  15.0)
\psbezier(  96.2,  15.0)(  96.0,  12.8)(  96.0,  11.0)(  95.9,  10.0)
\psset{linecolor=asgreen}
\psbezier(  65.7,  10.0)(  65.7,  11.1)(  65.7,  13.2)(  65.6,  15.0)
\psbezier(  65.6,  15.0)(  65.6,  16.9)(  65.6,  18.4)(  65.5,  19.9)
\psbezier(  65.5,  19.9)(  65.5,  21.6)(  65.4,  23.1)(  65.3,  24.9)
\psbezier(  65.3,  24.9)(  65.2,  26.6)(  65.0,  28.1)(  64.7,  29.8)
\psbezier(  64.7,  29.8)(  64.6,  30.8)(  64.3,  32.6)(  63.7,  34.6)
\psbezier(  63.7,  34.6)(  63.2,  36.3)(  62.6,  37.7)(  61.9,  39.2)
\psbezier(  61.9,  39.2)(  60.9,  41.2)(  59.9,  42.7)(  59.4,  43.5)
\psbezier(  59.4,  43.5)(  58.4,  44.9)(  57.4,  46.1)(  56.2,  47.3)
\psbezier(  56.2,  47.3)(  55.0,  48.6)(  53.9,  49.6)(  52.6,  50.7)
\psbezier(  52.6,  50.7)(  50.9,  52.1)(  49.4,  53.1)(  48.6,  53.6)
\psbezier(  48.6,  53.6)(  47.1,  54.5)(  45.7,  55.2)(  44.2,  55.9)
\psbezier(  44.2,  55.9)(  42.2,  56.7)(  40.5,  57.3)(  39.5,  57.6)
\psbezier(  39.5,  57.6)(  37.5,  58.2)(  35.7,  58.6)(  34.7,  58.8)
\psbezier(  34.7,  58.8)(  32.6,  59.2)(  30.8,  59.4)(  29.8,  59.5)
\psbezier(  29.8,  59.5)(  28.1,  59.7)(  26.6,  59.8)(  24.9,  59.9)
\psbezier(  24.9,  59.9)(  23.1,  60.0)(  21.6,  60.0)(  19.9,  60.0)
\psbezier(  19.9,  60.0)(  18.4,  60.0)(  16.9,  60.1)(  15.0,  60.1)
\psbezier(  15.0,  60.1)(  12.6,  60.1)(  12.3,  60.1)(  10.0,  60.1)
\psbezier(  96.6,  10.0)(  95.4,  11.8)(  96.0,  10.9)(  93.9,  14.0)
\psbezier(  93.9,  14.0)(  93.1,  15.1)(  92.5,  16.1)(  91.1,  17.8)
\psbezier(  91.1,  17.8)(  89.8,  19.6)(  89.1,  20.5)(  88.2,  21.6)
\psbezier(  88.2,  21.6)(  87.5,  22.4)(  86.2,  23.9)(  85.0,  25.2)
\psbezier(  85.0,  25.2)(  84.4,  25.8)(  83.2,  27.1)(  81.6,  28.5)
\psbezier(  81.6,  28.5)(  80.4,  29.6)(  79.3,  30.5)(  77.9,  31.4)
\psbezier(  77.9,  31.4)(  77.0,  32.0)(  75.3,  33.1)(  73.6,  33.6)
\psbezier(  73.6,  33.6)(  71.8,  34.1)(  70.3,  34.0)(  69.1,  33.1)
\psbezier(  69.1,  33.1)(  67.7,  32.1)(  67.1,  30.0)(  66.8,  29.0)
\psbezier(  66.8,  29.0)(  66.4,  27.3)(  66.2,  25.3)(  66.1,  24.3)
\psbezier(  66.1,  24.3)(  66.0,  22.6)(  65.9,  20.6)(  65.9,  19.5)
\psbezier(  65.9,  19.5)(  65.8,  18.1)(  65.8,  16.7)(  65.8,  14.8)
\psbezier(  65.8,  14.8)(  65.8,  12.6)(  65.8,  11.4)(  65.7,  10.0)
\psset{linecolor=ascyan}
\psbezier(  74.1,  10.0)(  73.7,  12.2)(  74.0,  11.0)(  73.2,  14.7)
\psbezier(  73.2,  14.7)(  72.8,  16.6)(  72.4,  18.0)(  72.0,  19.4)
\psbezier(  72.0,  19.4)(  71.5,  21.3)(  71.1,  22.6)(  70.6,  24.0)
\psbezier(  70.6,  24.0)(  70.3,  25.0)(  69.7,  26.6)(  68.9,  28.6)
\psbezier(  68.9,  28.6)(  68.2,  30.1)(  67.6,  31.4)(  66.9,  32.9)
\psbezier(  66.9,  32.9)(  66.1,  34.4)(  65.4,  35.7)(  64.5,  37.2)
\psbezier(  64.5,  37.2)(  63.7,  38.6)(  62.9,  39.8)(  61.9,  41.2)
\psbezier(  61.9,  41.2)(  61.3,  42.0)(  60.2,  43.4)(  58.9,  45.0)
\psbezier(  58.9,  45.0)(  57.7,  46.2)(  56.7,  47.3)(  55.5,  48.4)
\psbezier(  55.5,  48.4)(  54.8,  49.1)(  53.4,  50.3)(  51.8,  51.5)
\psbezier(  51.8,  51.5)(  51.0,  52.1)(  49.5,  53.1)(  47.7,  54.1)
\psbezier(  47.7,  54.1)(  46.2,  54.9)(  44.9,  55.6)(  43.4,  56.2)
\psbezier(  43.4,  56.2)(  41.4,  57.0)(  39.7,  57.5)(  38.8,  57.8)
\psbezier(  38.8,  57.8)(  36.8,  58.4)(  35.0,  58.7)(  34.1,  58.9)
\psbezier(  34.1,  58.9)(  32.0,  59.2)(  30.3,  59.4)(  29.3,  59.5)
\psbezier(  29.3,  59.5)(  27.6,  59.7)(  26.2,  59.8)(  24.5,  59.8)
\psbezier(  24.5,  59.8)(  23.0,  59.9)(  21.6,  59.9)(  19.7,  60.0)
\psbezier(  19.7,  60.0)(  18.2,  60.0)(  16.8,  60.0)(  14.8,  60.0)
\psbezier(  14.8,  60.0)(  12.6,  60.0)(  12.2,  60.0)(  10.0,  60.0)
\psset{linecolor=asblue}
\psbezier(  75.0,  10.0)(  74.5,  12.1)(  74.9,  11.1)(  74.0,  14.7)
\psbezier(  74.0,  14.7)(  73.6,  16.2)(  73.3,  17.5)(  72.7,  19.4)
\psbezier(  72.7,  19.4)(  72.4,  20.3)(  71.8,  22.0)(  71.0,  23.9)
\psbezier(  71.0,  23.9)(  70.2,  26.0)(  69.6,  27.1)(  69.0,  28.3)
\psbezier(  69.0,  28.3)(  68.6,  29.3)(  67.7,  31.1)(  66.8,  32.6)
\psbezier(  66.8,  32.6)(  65.8,  34.4)(  64.8,  36.0)(  64.3,  36.8)
\psbezier(  64.3,  36.8)(  63.4,  38.2)(  62.6,  39.4)(  61.6,  40.7)
\psbezier(  61.6,  40.7)(  60.6,  42.1)(  59.6,  43.2)(  58.5,  44.5)
\psbezier(  58.5,  44.5)(  57.9,  45.2)(  56.7,  46.6)(  55.2,  48.0)
\psbezier(  55.2,  48.0)(  54.5,  48.7)(  53.2,  49.9)(  51.6,  51.2)
\psbezier(  51.6,  51.2)(  50.8,  51.8)(  49.4,  52.9)(  47.6,  54.0)
\psbezier(  47.6,  54.0)(  46.2,  54.8)(  44.9,  55.5)(  43.4,  56.2)
\psbezier(  43.4,  56.2)(  41.5,  57.1)(  39.7,  57.6)(  38.8,  57.9)
\psbezier(  38.8,  57.9)(  36.8,  58.5)(  35.1,  58.8)(  34.1,  59.0)
\psbezier(  34.1,  59.0)(  32.1,  59.3)(  30.3,  59.5)(  29.3,  59.6)
\psbezier(  29.3,  59.6)(  27.6,  59.7)(  26.2,  59.8)(  24.5,  59.9)
\psbezier(  24.5,  59.9)(  23.0,  59.9)(  21.6,  59.9)(  19.7,  60.0)
\psbezier(  19.7,  60.0)(  18.2,  60.0)(  16.8,  60.0)(  14.8,  60.0)
\psbezier(  14.8,  60.0)(  12.6,  60.0)(  12.3,  60.0)(  10.0,  60.0)
\put(85,36){
\psline[linecolor=asblue](2,21)(8,21)
\put(10,21){\makebox(0,0)[l]{\small \sf $\chi = 0.411, \eta = 0.23$}}
\psline[linecolor=ascyan](2,16)(8,16)
\put(10,16){\makebox(0,0)[l]{\small \sf Blackman-Harris, $N = 6$}}
\psline[linecolor=asgreen](2,11)(8,11)
\put(10,11){\makebox(0,0)[l]{\small \sf $\chi = 0.414, \eta = 0.61$}}
\psline[linecolor=asorange](2,6)(8,6)
\put(10,6){\makebox(0,0)[l]{\small \sf Lanczos, $M = 2$}}
}
\end{pspicture}

\caption{Fourier transforms of some commonly used kernels, and of their approximation with the proposed parametrized
kernels. First set.}
\label{fgApproxGraphsA}
\end{figure}
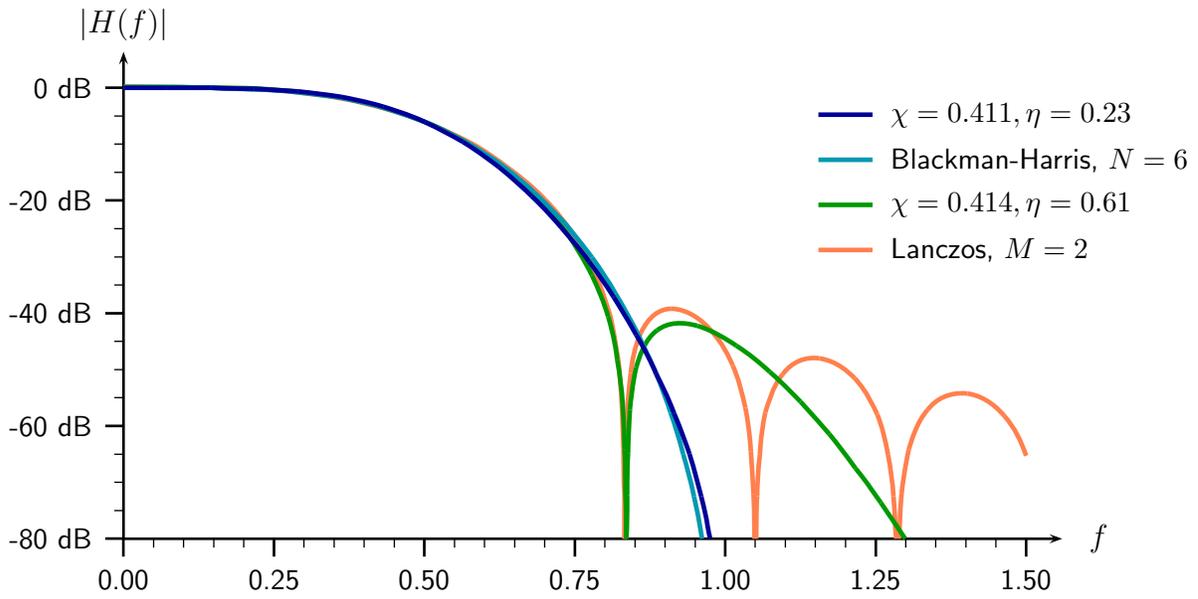

Many of the features of the new kernels can be observed by analyzing versions that have parameters chosen to closely
approximate kernels commonly used for interpolation. Table~\ref{tbParApprox} shows some sets of parameters that can
be used for these approximations, and Figures \ref{fgApproxGraphsA} and \ref{fgApproxGraphsA} show comparisons of the
corresponding Fourier transforms. In Figure~\ref{fgApproxGraphsA} we can observe that with $\chi=0.248$ and $\eta=0.48$
we have a response nearly identical to the Blackman-Harris ($N=6$) kernel~\cite{ Harris78,Lehman99}. Choosing
$\chi=0.163$ and $\eta=1.2$ produces a response very similar to the Lanczos kernel ($M=2$)~\cite[\S~3]{Glassner} up
to its first zero. After that, $H_s(f; \chi,\eta)$ produces a wider sidelobe with roughly the same height, but with
nearly monotonic decay, instead of several sidelobes.

%
\renewcommand{\arraystretch}{1.05}
%
\begin{table}
\centering
\caption{Parameters that approximate popular interpolation kernels.}
\label{tbParApprox}

\begin{small}
\vspace{1ex}
\begin{tabular}{|l||c|c|} \hline
 \bf Kernel & $\chi$ & $\eta$ \\ \hline \hline
 Lanczos, $M=2$ & 0.414 & 0.61 \\ \hline
 Lanczos, $M=3$ & 0.284 & 0.64 \\ \hline
 Lanczos, $M=4$ & 0.212 & 0.65 \\ \hline
 Lanczos, $M=5$ & 0.170 & 0.65 \\ \hline
 Blackman-Harris, $N=6$ & 0.411 & 0.23 \\ \hline
 Cubic B-Spline & 0.310 & 0 \\ \hline
 Mitchell-Netravali, $B=C=1/3$ & 0.550 & 0.32 \\ \hline
\end{tabular}
\end{small}
\end{table}

%

%
\begin{figure}[tbp]
\centering
\setlength{\unitlength}{1.2mm}
\psset{unit=1.2mm}

\input{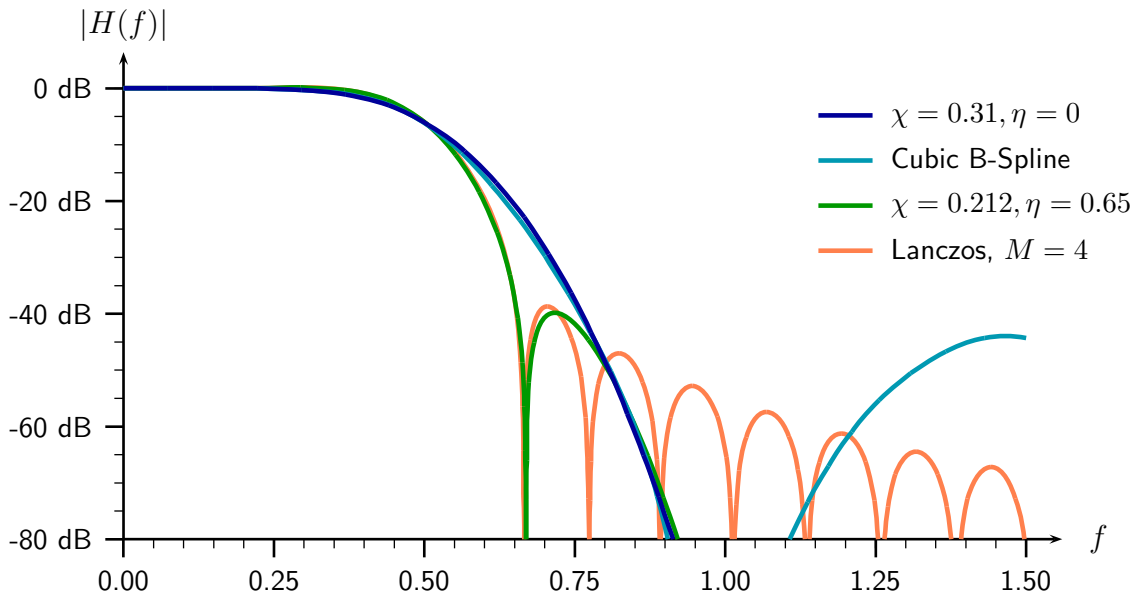}

\caption{Fourier transforms of some commonly used kernels, and of their approximation with the proposed parametrized
kernels. Second set.}
\label{fgApproxGraphsB}
\end{figure}

Figure~\ref{fgApproxGraphsB} shows a comparison with another Lanczos kernel, with similar results. In part, these
results are not very surprising, since the Blackman-Harris and Lanczos kernels are based on the sinc($t$) function.
However, they show that we can get remarkable control of the properties of $H_s(f; \chi,\eta)$ by changing only its two
parameters.

Furthermore, the comparison of $H_s(f; \chi,\eta)$ with the cubic b-spline kernel in Figure~\ref{fgApproxGraphsB}
shows that we can have very good approximations for other types of kernels too. One important difference, is that
$H_s(f; \chi,\eta)$ has no sidelobes above --80~dB when $f>1$, which means that it has better partition of unity when
used for downsampling.

We also tested the Lanczos (LZ4), Blackman-Harris (BH6), and cubic b-spline (CBS) kernels in the $2048 \times 2560$
image ``Bike'' (used in the JPEG2000 tests), chosen because it has many details and test patterns. First, we upsampled
all images by factor 1.7. In all cases, the images are visually indistinguishable, so we measured the differences
between images. The results, defined as PSNR in dB, are:

\begin{center}
\vspace{0.5ex}
\begin{small}
\begin{tabular}{|c||c|c|c|} \hline
 \bf Kernel & LZ4 & BH6 & CBS \\ \hline \hline
 LZ4-Approx. & 51.3 & 42.6 & 49.4 \\ \hline
 BH6-Approx. & 43.4 & 58.3 & 48.4 \\ \hline
 CBS-Approx. & 47.9 & 47.3 & 58.1 \\ \hline
\end{tabular}
\end{small}
\vspace{0.5ex}
\end{center}

The values in the main diagonal show that the images obtained with the kernels and their approximations are indeed very
close. The other values are also large, indicating that all those kernels produce good results, but the differences
are also clear.

In the second experiment we downsampled the image by factor 0.6, and obtained similar results

\begin{center}
\vspace{0.5ex}
\begin{small}
\begin{tabular}{|c||c|c|c|} \hline
 \bf Kernel  & LZ4 & BH6 & CBS \\ \hline \hline
 LZ4-Approx. & 51.4 & 38.9 & 45.6 \\ \hline
 BH6-Approx. & 39.9 & 56.1 & 45.2 \\ \hline
 CBS-Approx. & 45.2 & 44.0 & 57.0 \\ \hline
\end{tabular}
\end{small}
\end{center}

\begin{figure}[p]
\centering
\setlength{\unitlength}{1mm}
\psset{unit=1mm}

\begin{pspicture}(120,80)
\psset{linewidth=1pt}
\psline{->}(  10.0,  10.0)( 114.0,  10.0)
\multiput(  10.0,  10.0)(  20.000,   0.000){ 6}{\psline(   0.0,  -2.0)}
\multiput(  10.0,  10.0)(   4.000,   0.000){26}{\psline[linewidth=0.5pt](   0.0,  -1.0)}
\put(  10.0,   6.5){\makebox(0,0)[t]{\small \sf 0.0}}
\put(  30.0,   6.5){\makebox(0,0)[t]{\small \sf 0.1}}
\put(  50.0,   6.5){\makebox(0,0)[t]{\small \sf 0.2}}
\put(  70.0,   6.5){\makebox(0,0)[t]{\small \sf 0.3}}
\put(  90.0,   6.5){\makebox(0,0)[t]{\small \sf 0.4}}
\put( 110.0,   6.5){\makebox(0,0)[t]{\small \sf 0.5}}
\put( 117.0,  10.0){\makebox(0,0)[l]{\sf $t$}}
\psline{->}(  10.0,  10.0)(  10.0,  74.0)
\multiput(  10.0,  10.0)(   0.000,  15.000){ 5}{\psline(  -2.0,   0.0)}
\multiput(  10.0,  10.0)(   0.000,   3.000){21}{\psline[linewidth=0.5pt](  -1.0,   0.0)}
\put(   6.5,  10.0){\makebox(0,0)[r]{\small \sf -2 \%}}
\put(   6.5,  25.0){\makebox(0,0)[r]{\small \sf -1 \%}}
\put(   6.5,  40.0){\makebox(0,0)[r]{\small \sf  0 \%}}
\put(   6.5,  55.0){\makebox(0,0)[r]{\small \sf  1 \%}}
\put(   6.5,  70.0){\makebox(0,0)[r]{\small \sf  2 \%}}
\put(  10.0,  77.0){\makebox(0,0){\sf DC Response Error}}
\psset{linewidth=1.75pt,linecolor=asblue}
\psbezier(  10.0,  43.5)(  28.1,  43.3)(  35.0,  38.7)(  59.6,  38.7)
\psbezier(  59.6,  38.7)(  82.5,  39.0)(  70.6,  39.5)( 110.0,  39.3)
\psset{linecolor=ascyan}
\psbezier(  10.0,  47.2)(  27.7,  46.9)(  42.0,  40.4)(  59.3,  37.7)
\psbezier(  59.3,  37.7)(  76.9,  35.1)(  92.3,  37.6)( 110.0,  37.6)
\psset{linecolor=asgreen}
\psbezier(  10.0,  51.7)(  42.5,  51.6)(  68.1,  29.4)( 110.0,  29.7)
\psset{linecolor=asorange}
\psbezier(  10.0,  58.4)(  26.9,  58.1)(  41.3,  50.7)(  59.8,  40.1)
\psbezier(  59.8,  40.1)(  78.4,  29.5)(  92.9,  21.7)( 110.0,  21.8)
\psset{linecolor=asred}
\psbezier(  10.0,  40.0)(  56.6,  40.0)(  63.4,  40.0)( 110.0,  40.0)
\put(65,52){
\psline[linecolor=asblue](2,26)(8,26)
\put(10,26){\makebox(0,0)[l]{\small \sf Cubic B-spline, $\beta = 0.85$}}
\psline[linecolor=ascyan](2,21)(8,21)
\put(10,21){\makebox(0,0)[l]{\small \sf Cubic B-spline, $\beta = 0.80$}}
\psline[linecolor=asgreen](2,16)(8,16)
\put(10,16){\makebox(0,0)[l]{\small \sf Cubic B-spline, $\beta = 0.75$}}
\psline[linecolor=asorange](2,11)(8,11)
\put(10,11){\makebox(0,0)[l]{\small \sf Cubic B-spline, $\beta = 0.70$}}
\psline[linecolor=asred](2,6)(8,6)
\put(10,6){\makebox(0,0)[l]{\small \sf New kernel, $\chi = 0.31, \eta = 0$}}
}
\end{pspicture}

\caption{Comparison of DC response errors between the cubic b-spline kernel and its approximation. The error of the
new kernel has a nearly constant value of $-5\cdot10^{-6}$ for all values of $t$ and $\beta$ in the graph.}
\label{fgSplineDCError}
\end{figure}
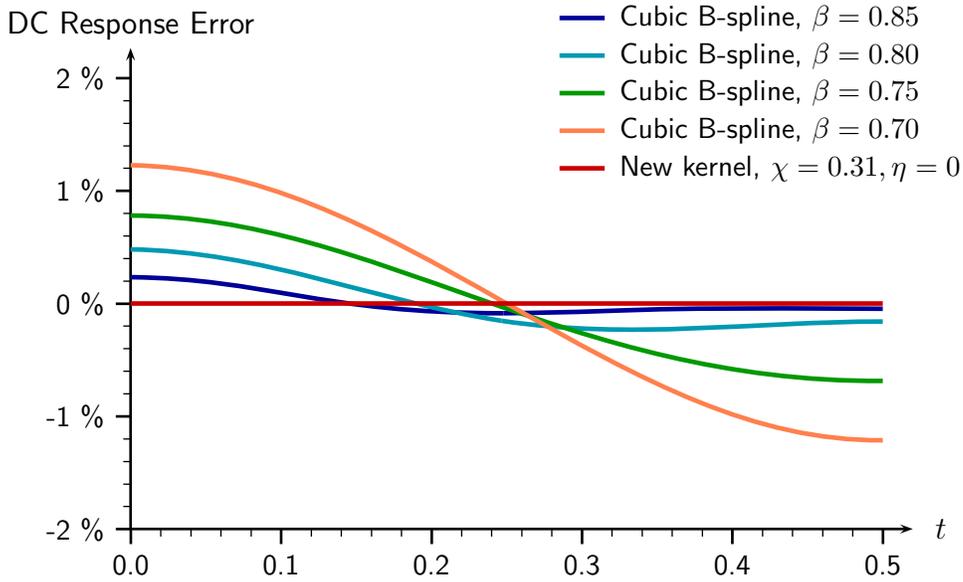

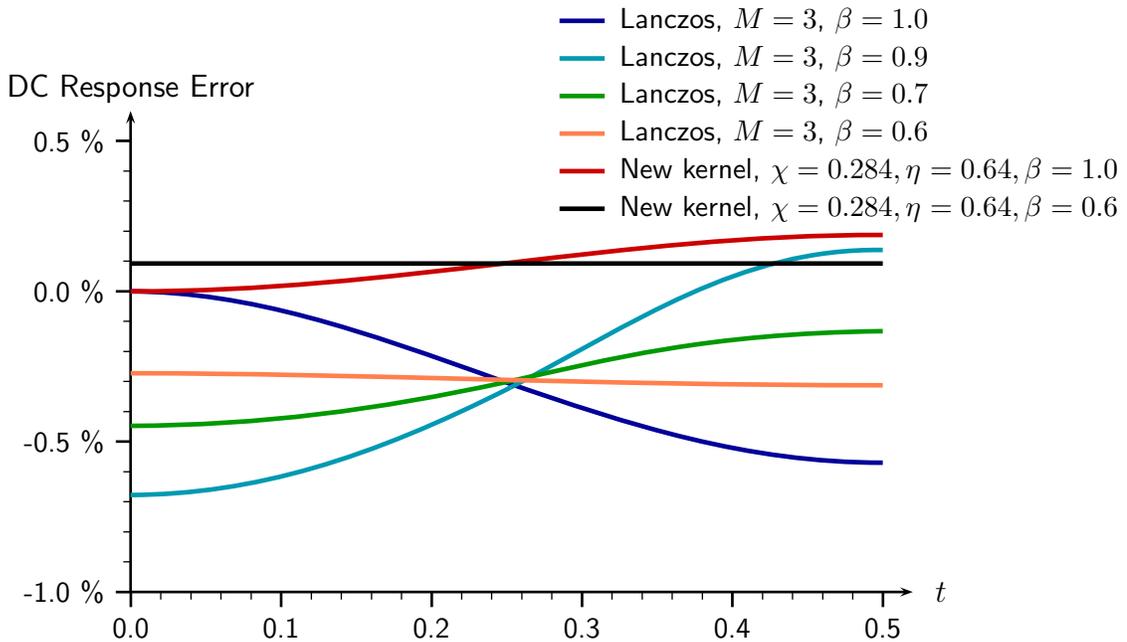
\begin{figure}[p]
\centering
\setlength{\unitlength}{1mm}
\psset{unit=1mm}

\begin{pspicture}(120,  80.0)
\psset{linewidth=1pt}
\psline{->}(  10.0,  10.0)( 114.0,  10.0)
\multiput(  10.0,  10.0)(  20.000,   0.000){ 6}{\psline(   0.0,  -2.0)}
\multiput(  10.0,  10.0)(   4.000,   0.000){26}{\psline[linewidth=0.5pt](   0.0,  -1.0)}
\put(  10.0,   6.5){\makebox(0,0)[t]{\small \sf 0.0}}
\put(  30.0,   6.5){\makebox(0,0)[t]{\small \sf 0.1}}
\put(  50.0,   6.5){\makebox(0,0)[t]{\small \sf 0.2}}
\put(  70.0,   6.5){\makebox(0,0)[t]{\small \sf 0.3}}
\put(  90.0,   6.5){\makebox(0,0)[t]{\small \sf 0.4}}
\put( 110.0,   6.5){\makebox(0,0)[t]{\small \sf 0.5}}
\put( 117.0,  10.0){\makebox(0,0)[l]{$t$}}
\psline{->}(  10.0,  10.0)(  10.0,  74.0)
\multiput(  10.0,  10.0)(   0.000,  20.000){ 4}{\psline(  -2.0,   0.0)}
\multiput(  10.0,  10.0)(   0.000,   4.000){16}{\psline[linewidth=0.5pt](  -1.0,   0.0)}
\put(   6.5,  10.0){\makebox(0,0)[r]{\small \sf -1.0 \%}}
\put(   6.5,  30.0){\makebox(0,0)[r]{\small \sf -0.5 \%}}
\put(   6.5,  50.0){\makebox(0,0)[r]{\small \sf  0.0 \%}}
\put(   6.5,  70.0){\makebox(0,0)[r]{\small \sf  0.5 \%}}
\put(  10.0,  77.0){\makebox(0,0){\sf DC Response Error}}
\psset{linewidth=1.75pt,linecolor=asblue}
\psbezier(  10.0,  50.0)(  26.0,  49.9)(  40.4,  44.9)(  59.6,  38.0)
\psbezier(  59.6,  38.0)(  76.6,  32.1)(  91.9,  27.2)( 110.0,  27.2)
\psset{linecolor=ascyan}
\psbezier(  10.0,  22.9)(  26.5,  23.0)(  41.7,  27.7)(  60.5,  37.2)
\psbezier(  60.5,  37.2)(  76.8,  45.6)(  90.9,  55.8)( 110.0,  55.5)
\psset{linecolor=asgreen}
\psbezier(  10.0,  32.1)(  27.5,  32.2)(  42.7,  34.2)(  59.8,  37.9)
\psbezier(  59.8,  37.9)(  73.9,  41.0)(  86.5,  44.4)( 110.0,  44.7)
\psset{linecolor=asorange}
\psbezier(  10.0,  39.1)(  41.0,  39.1)(  70.1,  37.5)( 110.0,  37.5)
\psset{linecolor=asred}
\psbezier(  10.0,  50.0)(  45.0,  49.9)(  75.0,  57.6)( 110.0,  57.5)
\psbezier[linecolor=black](  10.0,  53.7)(  56.6,  53.7)(  63.4,  53.7)( 110.0,  53.7)
\put(65,60){
\psline[linecolor=asblue](2,26)(8,26)
\put(10,26){\makebox(0,0)[l]{\small \sf Lanczos, $M=3$, $\beta = 1.0$}}
\psline[linecolor=ascyan](2,21)(8,21)
\put(10,21){\makebox(0,0)[l]{\small \sf Lanczos, $M=3$, $\beta = 0.9$}}
\psline[linecolor=asgreen](2,16)(8,16)
\put(10,16){\makebox(0,0)[l]{\small \sf Lanczos, $M=3$, $\beta = 0.7$}}
\psline[linecolor=asorange](2,11)(8,11)
\put(10,11){\makebox(0,0)[l]{\small \sf Lanczos, $M=3$, $\beta = 0.6$}}
\psline[linecolor=asred](2,6)(8,6)
\put(10,6){\makebox(0,0)[l]{\small \sf New kernel, $\chi = 0.284, \eta = 0.64, \beta = 1.0$}}
\psline[linecolor=black](2,1)(8,1)
\put(10,1){\makebox(0,0)[l]{\small \sf New kernel, $\chi = 0.284, \eta = 0.64, \beta = 0.6$}}
}
\end{pspicture}

\caption{Comparison of DC response errors between the Lanczos kernel and its approximation.}
\label{fgLanczosDCError}
\end{figure}

As explained in Section~\ref{ssInterpDown}, an advantage of the new kernels is that they present very fast decays on the
frequency response, which can make them (depending on the parameters) better for downsizing because of the smaller error
on the DC response. Figures \ref{fgSplineDCError} and \ref{fgLanczosDCError} show comparisons of the DC response error
for two kernel approximations. In Figure~\ref{fgSplineDCError} we observe that while the cubic b-spline---which has
ideal DC response when $\beta=1$---produces errors for other values of $\beta$. Its approximation, on the other hand,
has not sidelobe ($\eta=0$, cf. Figure~\ref{fgApproxGraphsB}), and consequently the very fast decay produces very small
DC response errors. In fact, the error has a nearly constant it $-5\cdot10^{-6}$ for all values of $0 \leq t \leq 1$
and $0.5 \leq \beta \leq 1]$. Figure~\ref{fgLanczosDCError} shows a similar comparison, with the Lanczos kernel. Note
that is this case the DC response of the approximation is similarly nearly independent of $t$ and $\beta$ (in fact,
$\beta=1$ is commonly the worst case). However, there are error because $\eta$ has a relatively large value.

%

%

\section{Conclusions}

We have shown that the proposed kernels for image interpolation and resizing can be easily designed, since their two
parameters provide direct control over the most important features, which are the width of the transition band, and
the sidelobe height. We also explain that the kernels naturally satisfy many desirable conditions. They yield exact
interpolation, both the functions and their Fourier transforms have very fast decays, and thus the same kernels can
produce good results for both interpolation and downsampling. We tested the flexibility of the design by presenting
sets of parameters that produce kernels that are very good approximations of kernels that are well known for their
properties and superior image quality. The differences between the original kernels and the approximations are evaluated
by analyzing the frequency response, and also measuring the difference between images resized with those kernels. In
conclusion, the new class of kernels provide a very convenient way to test different kernels in order to identify those
that produce the best image quality.




\end{document}